\documentclass[USenglish,oneside,twocolumn]{article}

\usepackage[utf8]{inputenc}

\usepackage[big]{dgruyter_NEW}

\usepackage{soul}
\usepackage{color,url}

\usepackage{enumitem,kantlipsum}

\usepackage{graphicx,url,subfigure}
\usepackage{adjustbox}

\usepackage{multirow}

\begin{document}


\title{ \huge The Road Not Taken: Re-thinking the Feasibility of Voice Calling Over Tor}
\runningtitle{The Road Not Taken: Re-thinking the Feasibility of Voice Calling Over Tor}

\author*[1]{Piyush Kumar Sharma}

\author[2]{Shashwat Chaudhary$^\dagger$}

\author[3]{Nikhil Hassija$^\dagger$}

\author[4]{Mukulika Maity}
  
\author[5]{Sambuddho Chakravarty}
 
\affil[1]{Indraprastha Institute of Information Technology (IIIT) Delhi, India E-mail: piyushs@iiitd.ac.in}

\affil[2]{IIIT Delhi, India E-mail: shashwat15091@iiitd.ac.in}

\affil[3]{IIIT Delhi, India E-mail: nikhil15065@iiitd.ac.in}

\affil[4]{IIIT Delhi, India E-mail: mukulika@iiitd.ac.in}

\affil[5]{IIIT Delhi, India E-mail: sambuddho@iiitd.ac.in

$^\dagger$ Both the authors have equal contribution.}

\begin{abstract}
{ 
Anonymous VoIP calls over the Internet  holds great significance for privacy-conscious users, whistle-blowers and political activists alike. 
Prior research deems popular anonymization systems like Tor unsuitable for providing requisite performance guarantees
that real-time applications like VoIP need. Their claims are backed by studies that may no longer be valid due to constant advancements in Tor. Moreover, we believe that these studies lacked the requisite diversity and comprehensiveness. 
Thus, conclusions from these studies, led them to propose novel and tailored solutions.
However, no such system is available for immediate use.
Additionally, operating such new systems would incur significant costs for recruiting users
and volunteered relays, to provide the necessary anonymity guarantees.\\
It thus becomes imperative that the exact performance of VoIP over Tor be quantified and analyzed, so that
the potential performance bottlenecks can be amended. 
The impact of interplay of the network performance attributes
(\emph{e.g.,} RTT, bandwidth, \emph{etc.}) on the perceived call quality thus also needs a fresh look to have a better understanding.
We thus conducted an extensive empirical study across various in-lab and real world scenarios to shed light on VoIP performance over Tor. In over 0.5 million measurements spanning 12 months, across seven countries and covering about 6650 Tor relays, we observed that \textit{Tor supports good voice quality (Perceptual Evaluation of Speech Quality (PESQ) \textgreater 3 and one-way delay \textless $400 ms$) in more than 85\% of
cases}. Further analysis indicates that in general for most Tor relays, the contentions due to cross-traffic were
low enough to support VoIP calls, that are anyways transmitted at low rates (\textless 120 Kbps).
Our findings are supported by concordant measurements using \texttt{iperf} that show more than the
adequate available bandwidth for most cases. 
Data published by the Tor Metrics also corroborates the same.
Hence, unlike prior efforts, our research reveals that Tor is suitable for supporting anonymous VoIP calls.
}
\end{abstract}



\maketitle







\section{Introduction}
Voice-over-IP (VoIP) applications that support traffic encryption are popular among
users who are concerned about their communication privacy. However, these applications do not safeguard the anonymity for Internet users residing in regimes, which may conduct surveillance (\emph{e.g.}, the ability of NSA to intercept conversations is widely known~\cite{verizon,nsaprismskype,nsasurveillance}). Moreover, there does not exist any functional VoIP based 
system that ensures communication privacy
and anonymity (along with real-time communication guarantees) required by privacy conscious Internet users, whistle-blowers \emph{etc.}

Popular privacy and anonymity preserving
systems like \emph{Tor}~\cite{syverson2004tor} hide the actual IP address of the
communication peers by routing their traffic via
a cascade of proxies. Since such systems reroute traffic via circuitous paths,
it is a widely held belief that they would incur intolerable delays for real-time applications like
VoIP. More importantly, while traditionally VoIP
relies on UDP traffic to ensure real-time guarantees, popular systems like Tor
are designed to transport
TCP traffic. 
Further, complex cryptographic handshakes, essential to the anonymity guarantees provided by such systems, 
may exacerbate the impact on performance, making them unsuitable for real-time applications.

Prior efforts on anonymous calling~\cite{danezis2010drac,le2015herd,heuser2017phonion,schatz2017reducing} unanimously agree with aforementioned shortcomings of Tor.
Some conclude this based on potentially biased results~\cite{rizal2014study}, involving relays only in Europe or only conducted a few hundred calls~\cite{heuser2017phonion}, lacking
diversity. 
Others, such as Le Blond \emph{et al.}~\cite{le2015herd}, did not conduct any study for measuring VoIP performance over Tor.
Overall, we believe that existing literature does not quantify
the actual interplay of network performance attributes (\emph{e.g.}
one-way delay (OWD), available bandwidth, \emph{etc.}) and how it 
impacts VoIP call quality over Tor. 
However, the belief that Tor is not competent enough to transport VoIP traffic is still prevalent~\cite{heuser2017phonion}.

Thus several novel VoIP architectures were proposed~\cite{danezis2010drac,le2015herd, heuser2017phonion} to tackle the shortcomings. 
Some of these architectures, like \textit{Phonion}~\cite{heuser2017phonion} and \emph{Herd}~\cite{le2015herd}, require a new volunteer-run network 
along with millions of active users
(like Tor) for providing anonymity guarantees. This requirement can be a major stumbling block. 
Moreover, in the absence of active users and significant
cross-traffic, comparable to that of Tor (that transports over 200 Gbit/s traffic per day~\cite{tor-metrics}), one cannot adjudge future anonymity and performance
assurances of these proposals. 
\textit{Most importantly, no such system is currently functional.}

Hence, in the absence of an existing anonymous voice calling system, 
we chose to determine the root cause(s) of poor performance
over Tor. After ameliorating them, one may expect to achieve
adequate voice call quality with anonymity guarantees equivalent to that provided by Tor. 
To that end, we began by conducting a pilot study 
where we made VoIP calls over the Tor network.
We also captured the network performance
attributes, in order to eventually identify how they
impact voice call quality. Following ITU guidelines and
prior proposals~\cite{heuser2017phonion}, we used 
one-way delay (OWD) and \emph{Perceptual Evaluation of Speech Quality} (PESQ)~\cite{pesq}
as metrics to judge VoIP call quality. 
The PESQ ascribes a value to judge the user-perceived audio quality in an automated manner.

We made 1000 consecutive calls
through individual Tor circuits\footnote{Using the standard
Tor client utility.}, with the
callee and caller machines under our control. Contrary to the prevalent notion of poor quality of calls via Tor, we observed good call quality, with average PESQ $\approx$ 3.8
and average OWD $\approx$ 280 ms. Overall, 85\% of the calls were acceptable (PESQ \textgreater 3 and OWD \textless 400 ms) as per ITU~\cite{itu2000g,itu2003recommendation}.

In the absence of substantial evidence of poor quality
calls, we  
went ahead and conducted an extensive longitudinal study involving 0.5 million voice calls over the Tor network, spread across 12 months.
These measurements not only involved varied in-lab and real-world scenarios with diverse Tor relays, VoIP applications, caller-callee locations but also a user study.
To our surprise, even then more than 85\% of voice calls had acceptable quality.


The PESQ metric varies inversely with distortions
in the perceived audio. Packet drop and jitters, which cause
such distortions, are an artifact of increased cross-traffic
contentions. \textit{A high PESQ score, accompanied with low overall OWD, in the majority of the cases thus
indicates low cross-traffic contentions, for VoIP calls.
Moreover, VoIP calls which are mostly transmitted at low bit-rates (\textless 120Kbps), 
incur less routing costs, and thus may not suffer much distortions. 
} 

Other network performance metrics like available bandwidth also varies 
with network cross-traffic volume.
Thus, concomitant available bandwidth during the call (over 1Mbps in 90\% cases\footnote{The bandwidth was measured by \texttt{iperf}.}), 
along with data published by Tor Metrics~\cite{tor-metrics}, confirms the reason for obtaining good results.

Overall, this first ever long-term study involving extensive evaluations of voice calls
over Tor, bore some interesting results and insights. We summarize them as follows:
\begin{enumerate}
\vspace{-3mm}
    \item In the vast majority of our experiments (\textgreater 85\%), involving voice
    calls over individual Tor circuits, we observed acceptable call quality, PESQ \textgreater 3 and OWD \textless 400 ms. This holds for a diverse set of scenarios:
    \begin{enumerate}
        \item Caller and callee spread across 7 countries (in three continents).
        \item Coverage of a total of 6650 Tor relays, 22 times more than previous studies \cite{rizal2014study}. 
        \item Popular VoIP apps such as Telegram and Skype.
        \item Both caller and/or callee using Tor circuits to establish calls with one another. 
        \item Different codecs and call duration.
        \item User study involving humans rating voice calls.
    \end{enumerate}
    \item Acceptable performance in the vast majority of cases was
    due to \emph{relatively low contention for VoIP
    in most Tor relays.}
     
\end{enumerate}

\vspace{-8mm}

\section{Background and Related work}
\vspace{-2mm}
In this section, we begin by describing the basic concepts behind VoIP calling and its evaluation metrics. Next, we briefly describe Tor, followed by a discussion on different types of anonymous calling scenarios. Finally, we describe related work in this domain.
\vspace{-5mm}
\subsection{Basic of VoIP}
\label{voip}
\vspace{-2mm}
Voice over IP enables real-time voice communication over IP networks. Any VoIP based system comprises of two primary
channels--one for control and signaling, and the other for transporting the actual encoded voice traffic. 
Session Initiation Protocol (SIP)~\cite{sip}, is an example of a popular VoIP signaling protocol. It includes
functionality like authenticating users, establishing and terminating calls, \emph{etc.}
SIP along with Session Description Protocol (SDP)~\cite{sdp} allows negotiating call related parameters
like codecs. Once the call is set-up, Realtime Transport Protocol (RTP)~\cite{rtp}, a UDP based
protocol, solely manages voice traffic. It adds sequence numbers to packets for in-order delivery, and buffers
them to minimize the impact of jitter.
\vspace{-5mm}

\subsection{QoS Metrics For Voice Calling}
\label{back:Qos}
\vspace{-2mm}
The following metrics are often used for measuring the quality of voice calls.

\begin{enumerate}
\vspace{-3mm}
    \item \textbf{Perceptual Evaluation of Speech Quality:} PESQ \cite{pesq} is the ITU specified and standardized metric for
    voice call quality evaluation. It estimates the user perceived call quality. PESQ computation requires both the source and recorded audio for comparison. The PESQ metric generates an objective score using its own algorithm which is then mapped to a subjective Mean Opinion Score (MOS). It is demonstrated that the score generated by PESQ highly correlates with the MOS score reported by actual users.
    The metric largely reflects distortions in recorded audio, which
    in-turn indicates the impact of network losses and jitters.
    
    It can assume values between 0 and 5, 5 being excellent, and 0 being poor and unusable.  However, in practice, values between 1 and 4.5 are observed more often. Following ITU specifications, in our tests, we considered calls with PESQ greater than three as acceptable.

    \item \textbf{Jitter and Packet Loss:} Jitter represents variations between subsequent packet arrivals. Such variations may arise due to packet reordering and losses. While VoIP users can endure minor losses, jitter may dramatically hamper the perceived call quality. Even non-permissible variations in either of these two metrics can sharply perturb PESQ, which relies on both of these metrics. 
    
    \item \textbf{One Way Delay (OWD):} OWD is the time duration between when voice packets are encoded at the sender and
    when they are successfully decoded at the receiver. According to the ITU specification~\cite{itu2003recommendation}, OWD should ideally be less than 150 ms. Further, OWD below 400 ms is considered permissible for international calls~\cite{itu2000g}. Hence, we chose the permissible limit of 400 ms to evaluate the performance of voice calls.

\end{enumerate}
 \vspace{-2mm} 
    Interestingly PESQ score, other than losses, only captures the impacts of jitters, which represents variations in OWD.
    It would suffer no perturbations if all the voice packets were uniformly delayed, without
    jitter or losses. However, for humans, such delays lead to time-shifted audio, eventually
    leading to perceptive confusion.
    Thus, in our evaluation study we considered both PESQ and OWD as metrics for estimating call quality.

\vspace{-5mm}
\subsection{The Onion Router (Tor)}
\vspace{-2mm}
Tor~\cite{syverson2004tor}, a widely-used low-latency anonymization
network, is built upon Chaum's MIXes~\cite{chaum1981untraceable}.
It allows it's users to communicate without revealing their
IP addresses. 
It consists of globally distributed volunteered hosts acting
as relays. Clients communicate to servers by proxying
their traffic via a cascade of three such relays, \emph{viz.}
the \emph{entry}, the \emph{middle}, and the \emph{exit} nodes.
The client encrypts the traffic using a three-layered encryption
scheme, each corresponding to the three relays, using
keys negotiated with each of them, respectively.
These encrypted packets are then forwarded via the three chosen relays.
Each of these relays de-crypts one layer of encryption and forwards
it to the next one in the cascade. Thus, no one ever, 
other than the client itself, knows \emph{the IP address of 
all the relays
and the server.} Each relay only knows about the previous
and next one in the cascade. The server only sees connections
arriving from the exit node, but knows nothing about the client.
By design Tor only supports TCP streams.

\vspace{-4mm}
\subsection{Types/Use Case Of anonymous Calling}
\label{anonuc}
\vspace{-2mm}

As already mentioned, anonymous calls are of great use to whistleblowers, activists, undercover reporters \emph{etc.}
However, one may ask the current
alternatives (used by such
groups) to conduct anonymous calls.
To the best of our knowledge, there are no publicly available alternatives. This is the reason that in the past decade, prior efforts~\cite{le2015herd,heuser2017phonion} attempted to design and build such systems. However, as discussed in Subsec.~\ref{relatedwork}, none of these systems are functional. Moreover, they would require the recruitment of a large number of volunteers to provide requisite anonymity guarantees similar to Tor~\cite{schatz2017reducing}. Thus, it becomes of prime importance to analyze an already available and actively maintained anonymity project Tor, to test the feasibility of conducting voice calls. 
However, there are secure messaging and end-to-end encrypted communication apps such as Signal and Telegram which are highly popular among privacy practitioners. Though these provide security against eavesdroppers, the centralized architecture of all such apps allows the central server to know the details such as who is calling to whom. Thus, even though such apps provide secure calls, they may not be fit to be used for anonymous calls.

We now describe the anonymous voice calling scenarios tested in our study.
\vspace{-3mm}
\begin{enumerate}

    \item \textbf{Caller (one-way) anonymity:} Here, the caller wants to achieve anonymity (by hiding  IP address) against an adversary that may monitor and/or filter its traffic. Such scenarios represent journalists and whistle-blowers who communicate sensitive information to other individuals or groups (\emph{e.g.}, news headquarters), while evading the adversary. 
    
    \item \textbf{Caller-Callee (two-way) anonymity:} Here, both parties want to achieve anonymity. Two individuals who both wish to communicate covertly, while remaining anonymous to their respective adversaries, may use such setups.
\end{enumerate}
\vspace{-3mm}
Further, these setups and their use cases are discussed in detail in Sec.~\ref{results}.
\vspace{-6mm}

\subsection{Prior Efforts}
\vspace{-3mm}
\label{relatedwork}

\begin{figure*}
\centering
\includegraphics[scale=0.6]{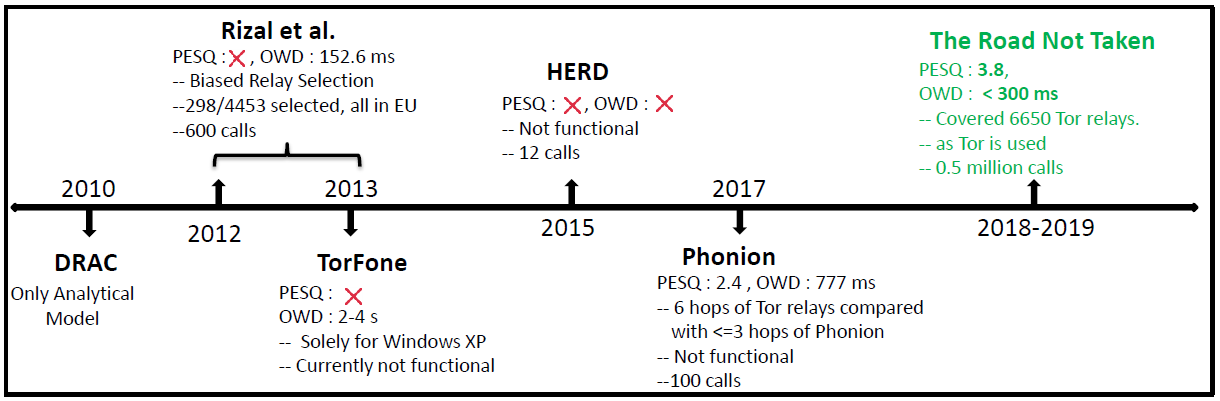}
\vspace{-2mm}
\caption{Existing literature on anonymous voice calling highlighting their inconsistencies
and incompleteness.
}
\label{timeline}
\vspace{-4mm}
\end{figure*}

There exists scant literature on performance evaluation of voice calls over Tor (ref Fig.~\ref{timeline}). The only attempts made were by Rizal \textit{et al.}~\cite{rizal2014study} and Heuser \textit{et al.}~\cite{heuser2017phonion}. These involved instantiating a few hundred calls through Tor for evaluating their quality. 

However, the aforementioned efforts were limited in scope.
\textit{E.g.}, rather than using PESQ (an established call quality metric), Rizal \textit{et al.} relied only on network parameters like OWD, jitter, and packet loss as evaluation criteria.
Also, this study involved Tor relays only in Europe (with only 298 out of the available 4453) for their evaluation, and may not be representative of VoIP performance, for the complete Tor network.
Based on these preliminary studies, prior efforts~\cite{le2015herd,heuser2017phonion,danezis2010drac} 
proposed novel architectures (ignoring Tor) for providing anonymous voice calls. 
Next, we describe all such efforts.

Inspired by Chaum's mixes~\cite{chaum1981untraceable}, Pfitzman \textit{et al.}~\cite{pfitzmann1991isdn} proposed ISDN mixes for anonymous voice communication.
Authors proposed two simplex Mix channels, one for the sender and  the other for the recipient, thereby enabling full-duplex anonymous communication over the telephony network.

Drac~\cite{danezis2010drac} by Danezis \emph{et al.}  involves an architecture for anonymous low latency voice communication using social networks as relays to route traffic. The system, while providing anonymity to a particular user, relies on using
the identity of other users in its social circle as alibis. 
However, it was an analytical model with no functional deployment.

\emph{Torphone}~\cite{torfone}, a system designed over Tor, was an extension to send VoIP traffic via Tor. It reported having achieved an OWD
of 2--4 s, which is unsuitable for acceptable call quality.
However, it is presently non-functional (and was last used on Windows XP).

In 2015, Le Blond \emph{et al.}~\cite{le2015herd}, proposed a novel architecture, \emph{Herd},
to prevent global passive/active adversaries attempts to
de-anonymize users, based on call metadata (correlating start and end times of a call). Herd relies on a set of dedicated mixes that relay VoIP traffic to mixes and endpoints while hiding any distinct traffic patterns. The mixes are also known as \textit{zones}, and a user can select available trustworthy zones to route its call. 
They believed prior results on evaluating the performance of VoIP over Tor (published
back in 2008~\cite{panchenko2008performance}), and concluded the RTT to be high (2--4 s), deeming Tor unsuitable for VoIP. However, they refrained from conducting a fresh study to gauge the (then) recent performance of Tor. 
Additionally, they concluded that the calling entities on Tor could be easily correlated, given the start and end times of a call. They validated this claim by analyzing the call records of a service provider. However, obtaining such data for calls conducted via Tor might not be easy, as Tor is a globally distributed system, and it would require gathering data from geographically diverse ISPs. 
Moreover, they tested their prototype on only four
cloud hosts, performing just 12 calls, in the absence of active users and significant cross-traffic when 
compared to Tor.

Phonion~\cite{heuser2017phonion} is one of the recent anonymous VoIP systems. It is fundamentally similar to Tor, but  specifically developed for voice communication. It 
uses \textit{relays} (similar to Tor relay), \textit{relay services} and \textit{broker system} (similar to Tor directory authorities). Phonion attempts to anonymize call data records (CDRs) against different adversaries. In Phonion, the
calls are relayed via various service providers. 
This prevents a single provider to gather all the call records
for a particular call. 
The advantage of Phonion is that it works across 
different voice calling technologies --- VoIP, cellular
or PSTN (public switched telephone network).
Additionally, it requires Internet access only for an initial bootstrap phase after which, anonymous calls could be instantiated over carrier services. 
However, the authors of Phonion also side stepped using Tor.
Through a pilot study of 100 calls, they concluded Tor
to be unsuitable for transporting VoIP calls. 
The study involved comparing one, two and three-hop circuits of their prototype, with six-hop Tor circuits (two-way anonymity).
However, we show in Sec.~\ref{sec:results:scenario1} and Sec.~\ref{sec:results:scenario2} that regular three-hop Tor circuits
provide suitable performance for VoIP.
Usage of this setup and performing 100 calls might have led them to report an unacceptable average
OWD of 777ms for Tor circuits along with an average PESQ of about 2.4.
Moreover, the implementation of their system relied on \texttt{google voice}, a service only available in N.America, making the system unusable elsewhere. Lastly, it required the relay operator to pay an operational fee to telcos, which might serve as a stumbling block for recruiting relays.

Overall, a major hindrance in the wide-scale adoption (and availability) of such systems, is the recruitment
of new volunteered relays and users (which Tor already has in abundance). Schatz \emph{et al.}~\cite{schatz2017reducing} also highlight this issue.
Additionally, as previously mentioned, prior evaluations over Tor, provide very few insights as to how the interplay of network performance attributes affects the
voice call quality. Thus, in this paper, we try to fill in these gaps by providing a comprehensive picture of the call quality achieved when using Tor.

\noindent\textbf{Comparison with prior VoIP measurements over Tor:} 
As already mentioned, only two studies actually measured the performance of VoIP over Tor --- Rizal's study~\cite{rizal2014study} and Phonion~\cite{heuser2017phonion} by Heuser \emph{et al.} We now explain how our measurement study is methodologically different from them.

\noindent\textit{Rizal's study:} This study transported VoIP traffic over Tor by tunnelling it via VPN tunnels. The experiments conducted in this study, involved Tor relays hosted only in Europe. Thus, Rizal reported a reasonably low average OWD of $\approx 152ms$ likely due to the geographic proximity of relays. On the other hand, 
our study included relays from all parts of the globe and hence provided better coverage of the entire Tor network\footnote{This was achieved by using the standard Tor utility that helped create different circuits based on default Tor circuit selection algorithm.}.
Thus, our study measured an average OWD of $\approx 280ms$ likely due to the diversity of involved Tor relays.
Moreover, unlike Rizal's study, we used PESQ as an evaluation metric, which is an industry standard for measuring user perceived voice quality.

\noindent\textit{Phonion:} It utilized the Mumble VoIP software~\cite{mumble} to transport VoIP traffic via Tor. Authors of Phonion used PESQ as an evaluation metric. However, their study involved measuring performance only for two-way anonymity, and not the one-way anonymity. 
On the other hand, our study involved different scenarios corresponding
to both the cases \emph{viz.} one-way and two-way anonymity.
Hence, we involved comparatively more diverse and comprehensive set of measurement setups. 

Additionally, both the previous studies involved performing only a few hundred calls --- 100 in Phonion and 600 in Rizal. 
In contrast, we performed $\approx0.5$ million calls involving a diverse set of geographic locations, media codecs, \emph{etc.,} along with controlled experiments involving private Tor setups. 

Overall, these studies lacked the requisite comprehensiveness and a detailed analysis of different performance attributes that could provide deeper insights for VoIP performance over Tor. Thus, our study aimed to fill this research gap by conducting an extensive study with an intent to better understand the behavior of VoIP calls over Tor in varied setups and network conditions.

\noindent\textbf{Other Tor performance measurement studies :}
\noindent There exist abundant studies which measure the Tor networks' performance in terms of observed bandwidth and latency \cite{onionperf,sbws,jansenshadow,dingledine2009performance,snader2010improving,mani2018understanding,jansen2019point,perry2009torflow,snader2009eigenspeed,johnson2017peerflow}.
Few, like Shadow \cite{jansenshadow} and Chutney \cite{chutney}, developed simulators to analyze the performance of Tor. Others like Torflow~\cite{perry2009torflow},
EigenSpeed \cite{snader2009eigenspeed}, and Peerflow \cite{johnson2017peerflow} \emph{etc.,} focused on 
measuring relay bandwidth for calculating relay weights by either directly measuring relay bandwidth (Torflow) or by indirectly inferring it using statistics such as relays reporting the amount of data exchanged between them (Peerflow).
Measuring relay bandwidth is a crucial task in Tor, as it is used to assign weights to different relays that govern the selection probability of a relay in a Tor circuit and thus the amount of traffic the relay might serve.
Cangialosi \emph{et al.} \cite{cangialosi2015ting} specifically focused on measuring RTT between relay nodes.
Tor metrics~\cite{tormetrics} is another popular (and actively maintained) project that periodically measures various performance attributes of the Tor network \emph{e.g.,} circuit RTTs, bandwidth for downloading files of various sizes, \emph{etc.}
However, none of these existing projects, measure the perceptual VoIP quality and its associated network characteristics. In our study, we primarily measure the VoIP performance over Tor, along with network characteristics (like bandwidth and RTT) of millions of Tor circuits. The metrics, such as available bandwidth are essential in our study as they help to infer the potential reasons for good or bad performance while conducting VoIP calls. We often referred to the Tor metric results to further support our claims.

\vspace{-6mm}

\section{Measurement Approach}
\label{sec:approach}
\vspace{-2mm}
In this section, we describe our experimental setups and the approach taken for performing experiments to measure the quality of voice calls over Tor.

\vspace{-6mm}
\subsection{Experimental Setup}
\label{setup}
\vspace{-2mm}
Previous efforts acknowledge that transporting VoIP traffic anonymously over Tor is non-trivial,
as VoIP generally uses UDP and Tor only supports TCP.

\begin{figure*}[!htb]
\minipage{0.45\textwidth}
  \includegraphics[width=\linewidth]{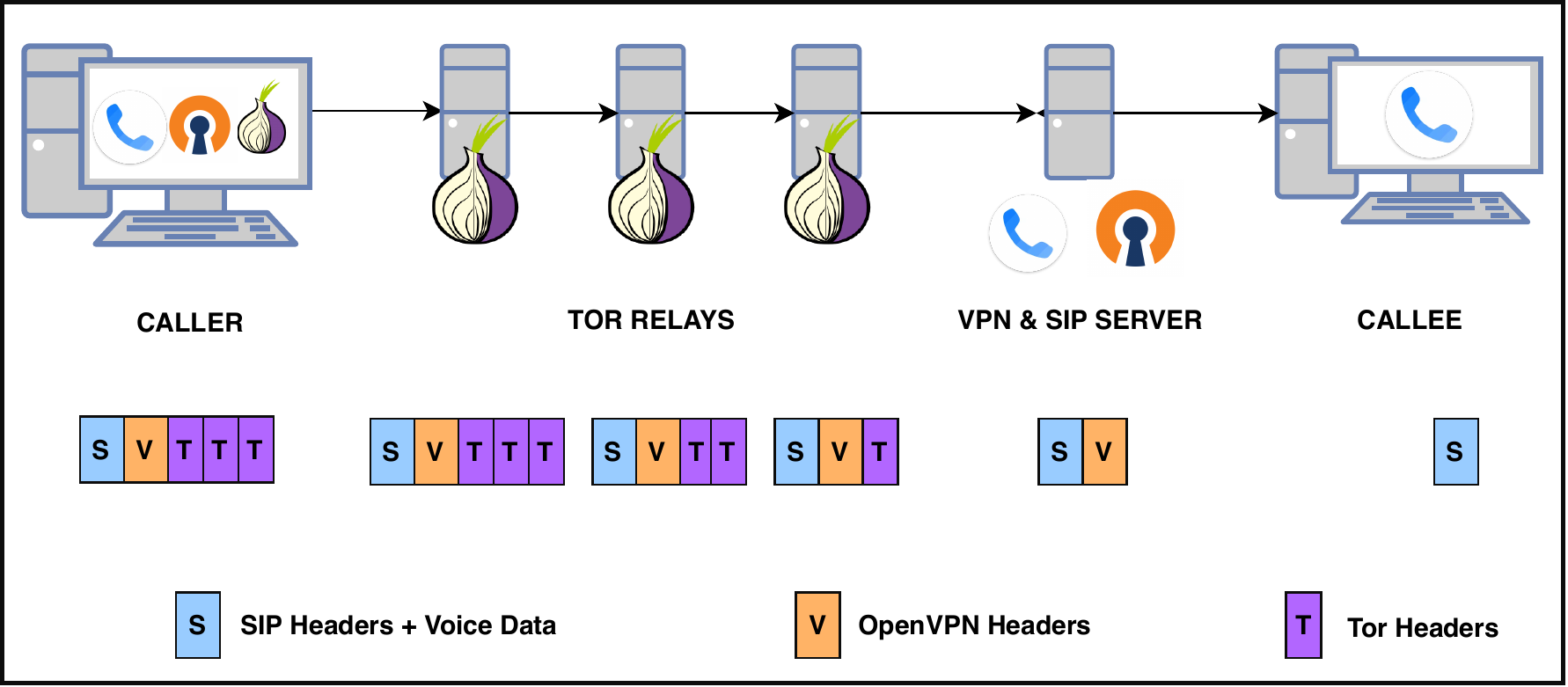}
  \caption{V-Tor: The caller establishes a VPN tunnel through Tor, so that all the UDP VoIP traffic traverses the Tor network. On reaching the VPN server, the traffic is sent to the SIP server, which initiates a voice call to the callee.}
  \label{VPN-setup}
\endminipage\hfill
\minipage{0.45\textwidth}
  \includegraphics[width=\linewidth]{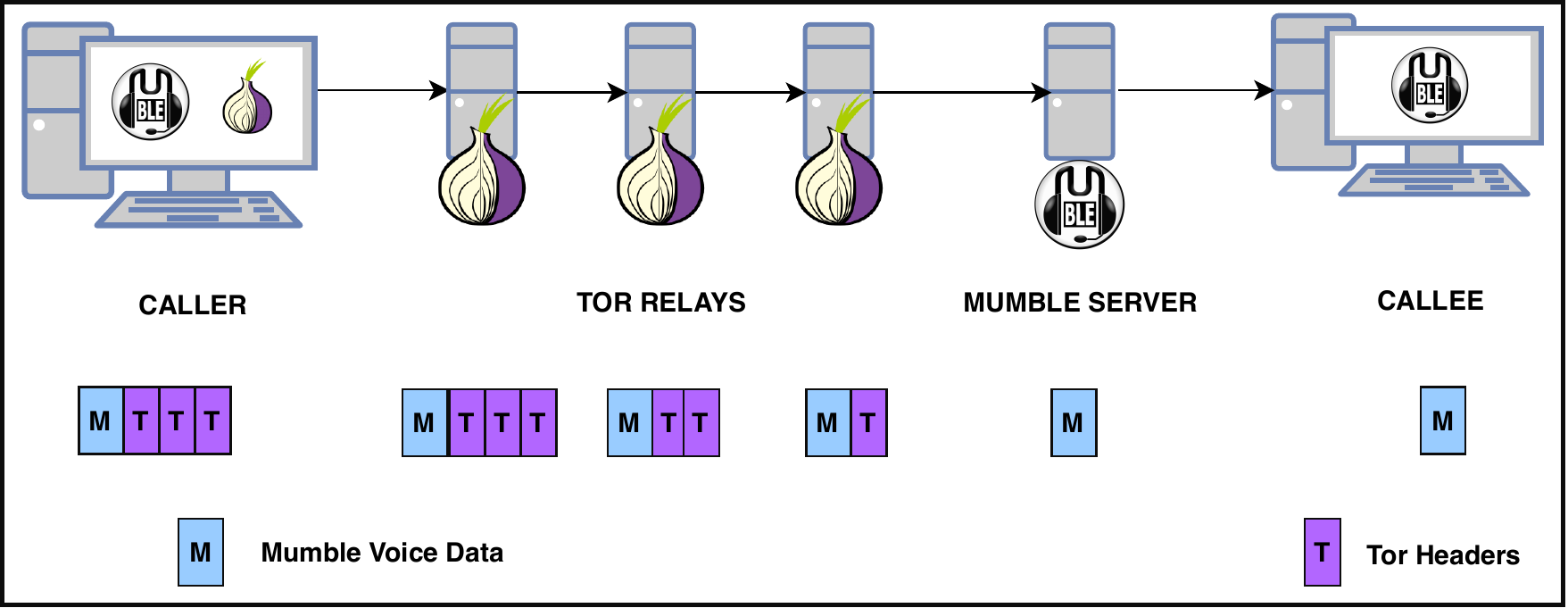}
  \caption{M-Tor: The caller configures the mumble client to work in TCP mode. Thereafter, mumble's traffic is sent over Tor eventually reaching the mumble server, which handles the call procedure between the caller and callee.}
  \label{mumble setup}
\endminipage
\vspace{-4mm}
\end{figure*}

There are two ways in which one can succeed in sending VoIP traffic over Tor. One way is to tunnel UDP packets inside TCP flows. The other way is to directly encode and send VoIP traffic in TCP packets. Thus, some previous studies~\cite{rizal2014study} utilized VPN tunnels to encapsulate and transfer VoIP packets, while others like Phonion~\cite{heuser2017phonion}, relied on Mumble~\cite{mumble} VoIP software to generate TCP packets and transfer them directly to Tor.
Similar to these studies, we tested VoIP performance using both the above approaches. These setups are now described in detail below:
\vspace{-2mm}
\begin{enumerate}
    \item \textbf{SIP client via VPN through Tor  \emph{(V-Tor)}:} 
    The V-Tor setup (ref. Fig.~\ref{VPN-setup}), 
    involves a SIP client (caller)
    connecting through a Tor circuit to a VPN server for establishing a TCP tunnel. 
    A step by step walkthrough of this setup is described below:
    
   
    \textbf{Step1} The caller runs the VPN and Tor client utilities. VPN client is configured to establish a VPN tunnel (to the VPN server) 
    over a Tor circuit, by forwarding VPN traffic to the Tor SOCKS interface.
    This ensured that all the traffic from the caller would reach the VPN server traversing the Tor network.
    
    \textbf{Step2} Client would initiate a VoIP call using a SIP utility. The VoIP traffic (generated from SIP) would reach the VPN server via Tor (explained in step 1).
    VPN server decapsulates SIP packets and forwards them to the SIP server.
    
    \textbf{Step3} SIP server then helps negotiate the call between the caller and the callee.
    
    

    
    \item  \textbf{Mumble with TCP mode over Tor (\emph{M-Tor}):} This setup does not rely on a VPN. Instead, it relies on using the Mumble client program (in TCP mode) for encoding voice traffic through TCP streams. These streams are transported via Tor to a Mumble server that mediates the voice call between the parties.
    
    
    
    

 \vspace{-2mm}
 \end{enumerate}

It must be noted that we performed experiments for both the V-Tor and M-Tor setups. However, we obtained similar results from both the setups. Thus, we present description of V-Tor experiments in the main body of the paper, and the details of M-Tor experiments in Appendix~\ref{app:mtor}.

 \vspace{-5mm}

\subsection{Overview of Experiments}
\vspace{-2mm}
We now describe the experiments performed to measure the VoIP call quality over Tor.
Our experiments were primarily conducted to identify the root cause of the
hitherto believed poor voice call quality of Tor. 

We began with a pilot study that involved conducting 1000 consecutive calls
over Tor, with the caller and callee under our control, but in 
different geo-locations. Our results, using V-Tor
(and M-Tor) setup, showed high call quality in a large fraction of the cases. Considering these results to be
potential outliers, compared to findings of previous authors,
we went ahead and conducted a comprehensive measurement study
spread across 12 months. 

We conducted two different sets of experiments --- 
(1) involving in-lab setups
and (2) involving circuits through the public Tor relays.


\noindent\textbf{In-lab experiments:} In the in-lab setup, our goal was to measure the performance of VoIP over V-Tor/M-Tor setups with competing cross-traffic entirely under our control. 
 For this, we setup a private network in our lab, consisting of Tor nodes along with client and server
(VPN, SIP, etc.) machines. This private network was deployed on real machines, with three of them serving as relays, while one of them was also serving as a directory authority. Other machines acted as clients and servers.
In these experiments, we measured performance attributes and established baseline values (\emph{e.g.,} bandwidth requirement) for a VoIP call. 
These experiments were performed over setups involving different combinations of VPN, Mumble, and Tor in the following manner:
\vspace{-3mm}
\begin{enumerate}
    \item \textit{Direct VoIP calls:} VoIP calls between the caller and callee were conducted without involving Tor and VPN.
    The caller and callee communicated directly using SIP or Mumble protocol. This helped us in observing the minimum bandwidth and delay requirements when no overhead was introduced (due to Tor or VPN).
 
    \item \textit{VoIP calls over VPN:}
    Caller and callee communicated using SIP protocol. However, calls were encapsulated through VPN connections, in order to measure the impact (if any) due to the overhead of running a VPN. 
    
    \item \textit{VoIP calls over Tor:}
    Encapsulating the calls through VPN connections (or Mumble) \emph{and} then transporting them via Tor circuits, to evaluate the impact of the overheads due to Tor.
    
\vspace{-2mm}
\end{enumerate}


Moreover, we performed some additional experiments to observe the impact of variation in background cross-traffic on VoIP calls. 
We observed the variation in performance when VoIP call(s) were made in the presence of heterogeneous background traffic (\emph{e.g.,} other VoIP calls and web traffic).
Such in-lab experiments may potentially present clues regarding the number of VoIP clients that could be simultaneously supported by the real Tor network.

\noindent\textbf{Internet based experiments over public Tor:} After performing various in-lab tests, we carried out multiple experiments involving public Tor relays, where we had no control over the background cross-traffic and network conditions. These experiments involved measuring performance across diverse scenarios (Tor relays, end-points, codecs, \emph{etc.}) with the
intention of studying the variation in performance under real-world conditions. 
More specifically, the experiments involved measuring the call quality by varying:
\vspace{-2mm}
\begin{itemize}

    \item \textbf{Tor circuits:} Involved measuring quality across a large number of circuits (6650 unique relays) created using the regular
    Tor client program.
   
    \item \textbf{Geo-location of communication peers:} Involved measuring quality by instantiating several
    calls, varying the location of the communication peers. 
   
    \item \textbf{Type of anonymity achieved:} Involved measuring quality while achieving one-way and two-way anonymity, in accordance with the use cases already described in Sec.~\ref{anonuc}.
    
    \item \textbf{Circuit lengths:} Involved  measuring quality over two-hop circuits. By default, Tor circuits are built using three relays.
    
    
    \item \textbf{Codecs:} Involved measuring quality by varying the the call codecs used. 
    
    \item \textbf{Call duration:} Involved measuring quality when call duration was varied.
    
    \item \textbf{Type of relays used:} Involved measuring quality when using bridges instead
    of public Tor relays.
    \vspace{-2mm}
\end{itemize}
    
We also measured the voice call quality for few popular voice calling apps such as \texttt{Telegram} \cite{tg} and \texttt{Skype} \cite{skype}, when used over Tor. Users may choose
to rely on using these already popular and familiar apps, instead of having
to setup V-Tor and M-Tor. 
Additionally, we also conducted a user study involving 20 participants who rated calls via Tor.
Most of experimental results are presented in the next section (Sec.~\ref{results}). A small fraction, addressing important concerns related to VoIP performance, \emph{e.g.}, impact of call codecs, Tor bridges, \emph{etc.} are presented in Sec.~\ref{sec:Discussion}.

\vspace{-4mm}
\subsection{Implementation Details}
\label{implementation}
\vspace{-2mm}

\noindent \textbf{\textit{Host configurations:}}
All machines of the in-lab experiments used Intel Core i5 8th gen CPUs with 8 GB of RAM. The hosts used in the experiments involving public Tor relays were hosted on Digital Ocean's cloud based infrastructure, distributed across seven countries. These machines were equipped with Intel Xeon 2.2 GHz single core CPUs and had 2 GB RAM. 
Since the latter experiments required initiating only a single call at a time, we did not require machines with high-end configurations.


\noindent \textbf{\textit{Tor configuration:}} In the basic configuration, all the caller and callee hosts had the latest Tor v0.3.9 installed. We used the Tor \texttt{stem} python library~\cite{stem}  
so as to ensure that only a single circuit was enabled at a time.


\noindent \textbf{\textit{Communication Peers:}} For V-Tor experiments, the end-point hosts were installed with OpenVPN~\cite{openvpn} v2.4.7. The VoIP traffic 
generated was encapsulated through OpenVPN client
and transported via Tor (by specifying the SOCKS port in the OpenVPN configuration). The clients used the python based SIP client, \texttt{pjsua}~\cite{pjsua}, a command-line softphone, to automate SIP calls. The module supported audio playout and recording.

For M-Tor experiments, the end-point hosts were installed with 
Mumble client program, configured to work in TCP mode. 
This enables directly sending voice traffic over TCP
streams. Further, like OpenVPN, the Mumble client also allows
transporting the voice traffic through 
Tor, by specifying the SOCKS port in its configuration.
\texttt{Torsocks}~\cite{torsocks} utility was used to transport \texttt{iperf}'s traffic over Tor. Since \texttt{torsocks} does not allow programs with root privileges, we relied on \texttt{socat}~\cite{socat} tunnels for transporting \texttt{ping}s.


\noindent \textit{\textbf{VPN/SIP Server:}} The V-Tor setup uses a machine configured to be a VPN as well as a SIP server. 
OpenVPN~\cite{openvpn} v 2.4.7 was configured to work as the VPN server. It forwarded the encapsulated SIP calls to the SIP server. 
Freeswitch PBX~\cite{freeswitch} was used as the SIP server for handling SIP calls. We used different configuration files that come with Freeswitch to make call extensions, route calls, select codecs, \emph{etc.}


\noindent \textit{\textbf{Popular VoIP apps:}} Among all the popular apps, Telegram is the only one that provides user APIs. Still, we 
require to overcome several challenges to measure the performance of
Telegram calls over Tor. 

Firstly, the API provides only instant messaging automation facility.
We thus mined the source code and discovered that it relies
on \texttt{libtgvoip}~\cite{libtgvoip} for making voice calls. We
used this library for automated calls. 
Secondly, the library uses UDP for transporting VoIP. We 
thus modified it to enforce transporting voice calls via TCP
streams. 
Thirdly, \texttt{libtgvoip} is configured, by default, to
play audio clips endlessly in a loop. We applied appropriate
modifications to control the playout duration, to suit
our calls. 
Fourthly, we also required modifying the library so as to
synchronize the call setup and termination events
with starting and stopping of voice recording. This
synchronization is required for accurately measuring
PESQ. 
Additionally, pyrogram~\cite{pyrogram} redirected Telegram traffic
to the Tor SOCKS port.

Other popular voice calling apps like Skype poses two challenges---\emph{viz.,} absence of resources like APIs or source codes to aid call automation, and ability to redirect traffic through Tor by configurable SOCKS interface. To overcome the first obstacle
we initially synced the caller and callee machines. Thereafter once the call was initiated, the caller plays out the audio clip using \texttt{mplayer}~\cite{mplayer}, while the callee
captured and recorded the call audio directly from the sound card using pactl~\cite{pactl} utility. 
To overcome the second challenge, we used OpenVPN to encapsulate
and redirect Skype call traffic to the Tor SOCKS interface.


\vspace{-4mm}    

\section{Measurement Results}
\label{results}
\vspace{-2mm}


In this section, we describe the experiments conducted to test the performance of anonymous calls over Tor, and their corresponding outcomes. 
We begin by enlisting some common steps we followed while conducting the experiments: 
\vspace{-2mm}
\begin{itemize}
    \item In all our experiments, a caller host played out an audio clip containing 30 s of human speech. It was encoded and transported, via a unique Tor circuit,
to the callee. The callee recorded the audio, which is later used for computing PESQ. 
 
    \item For every call, we recorded the network traffic through \texttt{pcap} files, and also measured various network performance attributes of the Tor circuit (through which the call is performed) like available bandwidth and RTT using \texttt{iperf} and \texttt{ping},
    respectively. The \texttt{ping} test was performed during a call as it is not a bandwidth intensive test. On the contrary, \texttt{iperf} (bandwidth intensive) tests were conducted after the completion of the call, so that it does not have any impact on the call quality.
    
    For the in-lab experiments also, we measured the stream bandwidth using \texttt{iperf}.
    
 
    \item PESQ score for every call was calculated by comparing the original (one played out at the caller) and recorded (at the callee) audio clips. Any score above three was considered good~\cite{itu2003recommendation}.

    \item One way delay was also calculated for the duration of the call. We used \texttt{ping} to calculate OWD.
    As per ITU guidelines for international calls~\cite{itu2000g}, the upper limit for OWD for acceptable voice call quality is $400$ ms.

\vspace{-2mm}  
\end{itemize}

In every experiment, the above steps were repeated for each call. Thereafter, we
analyzed the measurements and performance metrics across all these iterations. 

\vspace{-5mm}

\subsection{In-Lab Experiments}
\label{controlled}
\vspace{-2mm}

We performed these experiments, with an intent of measuring call performance under different testing
conditions, while fully controlling network link capabilities
and background cross-traffic. To establish the baselines, we created three test scenarios for the V-Tor setup. These involved:
(1) Direct SIP calls (2) SIP calls over VPN
tunnel (3) SIP calls through VPN over Tor (V-Tor). 

All these experiments followed the setup described in Fig.~\ref{VPN-setup}
For the scenarios where Tor was not used, the nodes between caller and callee (in Fig.~\ref{VPN-setup} 
) merely functioned as routers.
This was done to minimize any biases in performing experiments, by ensuring that the packets traverse the same number of hops\footnote{Additionally, removing these hops do not have any observable impact on our results. We kept them just to have uniformity among our experiments.}.


%

\noindent \textbf{Experiments using V-Tor setup:} We started by 
initiating VoIP calls over all the three setups and computed
their respective PESQ scores and OWD values.  The capacity of the link between the caller and callee was 100Mbps. The measured PESQ score averaged
across 100 individual samples was the same for all the three scenarios
\textit{i.e.}, 4.5. Whereas OWD was below 50 ms. This result established that for a single call, with no
competing cross-traffic, the overheads introduced by the VPN
and Tor had no significant impact on the call quality. We additionally
observed that the available bandwidth requirement for a single
call in all the three scenarios was no more than 120 Kbps (ref.
Tab.~\ref{tab:BandwidthVTor}). As expected, direct calls transmitted at
the lowest rates. Additional
overheads due to the headers introduced by VPN and Tor progressively increased
the bandwidth requirements. 

\begin{table}[h]
\vspace{-3mm}
\centering
\small
\begin{adjustbox}{max width=\textwidth}
\begin{tabular}{c c}
\hline
\textbf{Call category} & \textbf{Bandwidth (Kbps)} \\
\hline
Direct SIP call & 84 \\
SIP call via VPN & $\approx$ 108  \\
V-Tor & $\approx$ 120\\
\hline
\end{tabular}
\end{adjustbox}
\caption{Baseline bandwidth (in Kbps) requirement of VoIP in different scenarios.}
\label{tab:BandwidthVTor}
\vspace{-8mm}
\end{table}

Next, to understand the impact of cross-traffic on VoIP call quality, we initiated VoIP calls in the presence of cross-traffic.
The experiments were carried out for three link bandwidth
configuration---2 Mbps, 5 Mbps and 10 Mbps. Studying the performance under cross-traffic, for different link bandwidth would help us understand if the observed behavior is consistent or not. These experiments were specifically conducted for the VPN via Tor (V-Tor) setup.
Further, these lab experiments provided us insights on the number of calls that could potentially be made under varied network conditions on the real-Tor network.

Thus, we gradually introduced the cross-traffic by increasing the number of parallel file downloads
(using \texttt{wget}) from another client that shared the link with the caller. We made sure that the cross-traffic was in the direction of call, to ensure adequate cross-traffic contention.
We measured the degradation in the call quality, by computing
the average PESQ score, for every new parallel connection
introduced. Our findings are summarized in Tab.~\ref{tab:crossvtor}.

\begin{table}[h]
\vspace{-2mm}
\centering
\small
\begin{adjustbox}{max width=\columnwidth}
\begin{tabular}{c c c c l}
\hline
\textbf{Competing} & \textbf{Link} & \textbf{Available Bandwidth} & \textbf{Call} & \textbf{PESQ} \\
\textbf{Streams} & \textbf{Bandwidth} & \textbf{Per Stream} & \textbf{Requirement} & \textbf{Score} \\\hline
$<75$ & 10 Mbits & $>133$ Kbps & 120 Kbits & $>4.2$ \\\hline
80 & 10 Mbits & $125$ Kbits & 120 Kbits & $\approx 3.4 \downarrow $  \\\hline
$>85$ & 10 Mbits & $<117$ Kbits & 120 Kbits & $<2.3 \downarrow \downarrow$ \\\hline
\end{tabular}
\end{adjustbox}
\caption{Analysis of V-Tor under the presence of competing non-VoIP (web/file downloads) cross-traffic.}
\label{tab:crossvtor}
\vspace{-8mm}
\end{table}

We then performed experiments, where the background cross-traffic constituted of other VoIP calls. 
The cross-traffic was gradually increased such that it utilized the total link capacity from 5\%, to 10\%, and then all the way up to the point where the call under consideration received $<120$ Kbps of the total available bandwidth. At this point, we observed a sharp decline in PESQ for the call (\emph{i.e.,} $2.3$). This corresponds to unacceptable call quality.
Similar to the previous experiment, we performed this test for three different link capacities (2, 5 and 10 Mbps) for the V-Tor setup. The results of the 5 Mbps link bandwidth test are summarized in Tab.~\ref{tab:crossvtorvoip}. We obtained similar behavior for the other two bandwidth categories.

\begin{table}[h]
\vspace{-2mm}
\centering
\small
\begin{adjustbox}{max width=\columnwidth}
\begin{tabular}{c c c c l}
\hline
\textbf{Competing} & \textbf{Link} & \textbf{Available Bandwidth} & \textbf{Call} & \textbf{PESQ} \\
\textbf{VoIP Calls} & \textbf{Bandwidth} & \textbf{Per Call} & \textbf{Requirement} & \textbf{Score} \\\hline
$<35$ & 5 Mbits & $>145$ Kbps & 120 Kbits & $>4.2$ \\\hline
40-43 & 5 Mbits & $128$ Kbits & 120 Kbits & $\approx 3.3 \downarrow $  \\\hline
$>43$ & 5 Mbits & $<120$ Kbits & 120 Kbits & $<2.3 \downarrow \downarrow$ \\\hline
\end{tabular}
\end{adjustbox}
\caption{Analysis of V-Tor under the presence of competing VoIP cross-traffic.}
\label{tab:crossvtorvoip}
\vspace{-8mm}
\end{table}

The results indicate that when the contention on the shared link increases, the PESQ drops. The PESQ metric
is very sensitive to the impact of even minor network drops or delays. Even a small increase
in contention, \emph{e.g.} only five additional download streams reduce PESQ from 4.2 to 3.4 (ref. Tab.~\ref{tab:crossvtor}).
Corresponding to increased contentions, the available bandwidth for every stream (including VoIP)
drops. The constant bit-rate voice traffic of 120 Kbps suffers significant distortions that may, however,
have little impact on the non-VoIP flows.
This held when the cross-traffic was non-VoIP as well as when it was VoIP.

These in-lab measurements indicate that, 
a client should be able to conduct good quality calls, if the constructed circuit provides $>$ 120 Kbps bandwidth.
This should hold true on the real Tor network as well.
\emph{E.g.,} if a circuit has an available bandwidth of about $1.2$ Mbps, then it is capable to simultaneously support a maximum of 10 VoIP clients (with acceptable call quality).




\vspace{-5mm}
\subsection{Experiments involving public Tor relays}
\label{publictor}
\vspace{-2mm}
Having obtained good performance in in-lab tests, we went ahead to evaluate the performance of voice calls
over public Tor network. We expected the results to vary significantly due to the dynamic nature of 
competing cross-traffic and network conditions over the Internet.

We began with our pilot study that involved a client host (caller), positioned in our university, establishing 
VoIP call to a cloud hosted peer (callee). The call traffic was transported via Tor. We sequentially started 100
calls, each transported via a different Tor circuit, and measured the call quality by computing
PESQ score. To our surprise, for both the setups we observed an average PESQ of 3.8
and an acceptable OWD of 280 ms.
Even after 1000 calls, each transported via
a freshly created Tor circuit, we observed very similar performance measures (PESQ $\approx$ 3.86 and
OWD $\approx$ 273 ms).

However, one may argue that our positive results might have been a small anomalous fraction.
These may have been different from the bulk of
poor outcomes that may have led others to deem
Tor as unfit for VoIP. 
In order to test that this was not a fluke,
we conducted
a longitudinal experimental study covering diverse scenarios. 
The experiments are described below.


\vspace{-5mm}
\subsubsection{\textbf{Caller anonymity: Co-located voice server and callee (Scenario I)}}
\label{sec:results:scenario1}
\vspace{-2mm}
We begin with the fundamental scenario where a caller is positioned in
a censored network and wishes to call someone who is beyond the censor's
control. The caller makes calls to the callee, through
the public Tor network, using setups similar to one shown in Fig.~\ref{VPN-setup} and ~\ref{mumble setup}.
It is assumed that callee runs a publicly accessible
VPN server (for V-Tor)
on its host. This minimizes the overhead from the callee side.

 In our experiments, we chose seven individual cloud machines
as callers, and three other as callees. Each of these was selected from 
Europe, N. America, and Asia. 
For every caller--callee pair we made 1000 calls using the V-Tor 
setup. In each case, we measured the PESQ and OWD. A total of 
42000 calls were made. The average PESQ and OWD across all measurement was 3.88 and 217 ms, respectively.

\begin{figure}[h]
\centering
\vspace{-8mm}
\includegraphics[scale=0.26]{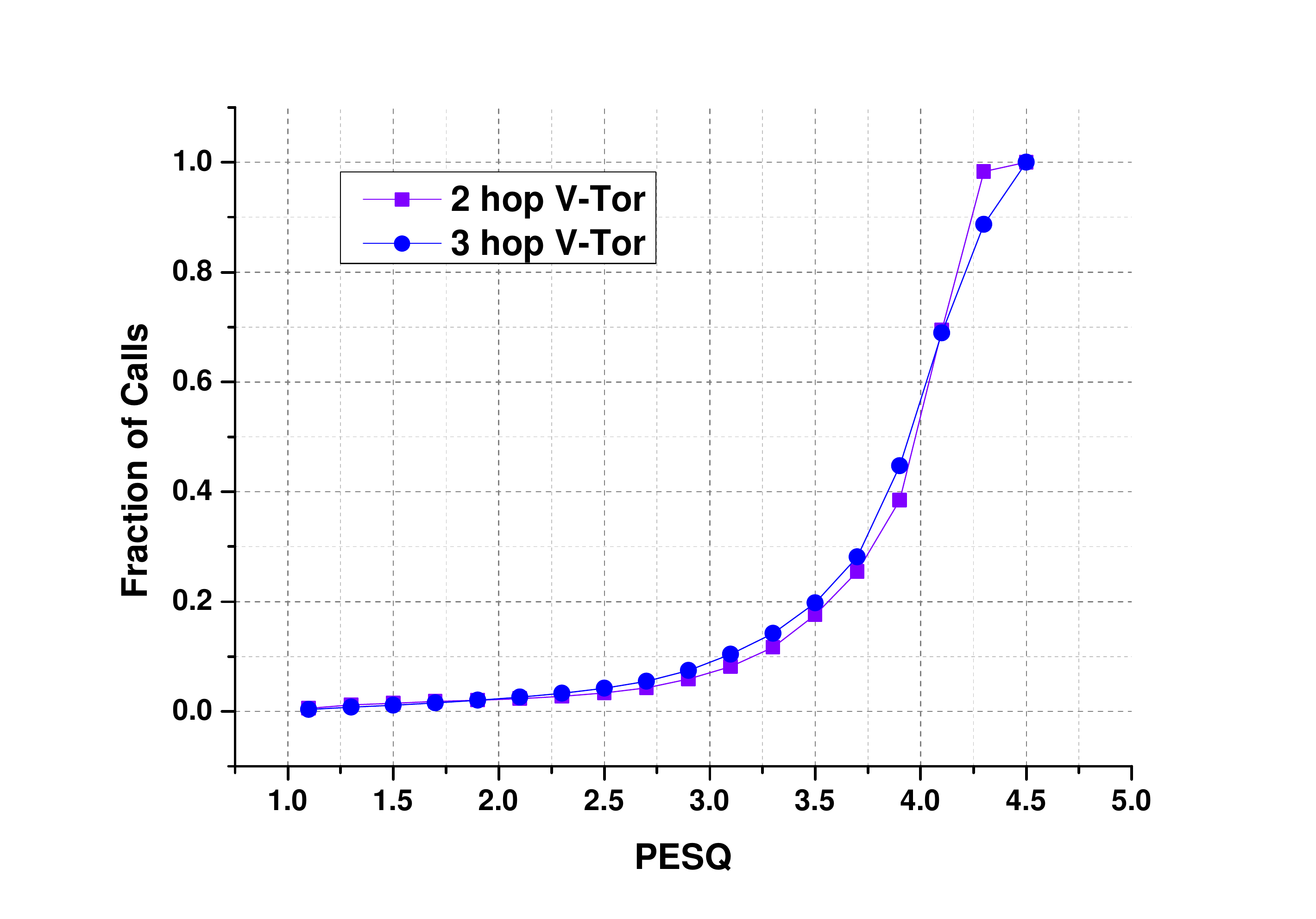}
\vspace{-5mm}
\caption{V-Tor: CDF of PESQ for Caller Anonymity when server is co-located with callee (Scenario I).}
\label{vctos}
\vspace{-6mm}
\end{figure}

By default, Tor circuits are three hops long. To reduce the potential
impact of hop length on performance, we repeated the said experiments
by making calls over two-hop circuits.
This did not led to
any significant impact on the PESQ (3.90). However, as expected, the one-way delay reduced to 205 ms. The CDF of PESQ scores obtained for these
experiments are shown in Fig.~\ref{vctos} 
~\footnote{We show graphs for the experiments conducted via V-Tor setup in the main paper. All the M-Tor setup results are moved to Appendix~\ref{app:mtor} for reference. Also, the overall results do not vary much.}
Results clearly show PESQ \textgreater 3 in over 93\% of the calls. 

\vspace{-6mm}
\subsubsection{\textbf{Caller anonymity: separate VPN/voice server\\
(Scenario II)}}
\label{sec:results:scenario2}
\vspace{-2mm}

There are, however, certain limitations of Scenario 1. 
Firstly,
the setup requires a publicly accessible VPN 
server, which
may be infeasible when the callee is behind a NAT. 
Secondly, there may be cases where multiple whistleblowers or covert reporters (\emph{i.e.}, several callers), 
communicate to callees working for a common organization. In such cases, having
a commonly shared VPN/SIP server, with high availability, supporting
features like voicemail, removes the need for the callee to be always online.
Thirdly, it reduces the hassle for every callee to
port VoIP server to different platforms.

Therefore we considered an alternative setup where the VPN / SIP
server
and callee were not hosted on the same host. They were distributed
among seven different cloud hosts, positioned across Europe, N. America and Asia. This separation may incur higher OWD between the communication
peers, due to the intervening network between the
VPN/SIP server and the callee, thus impacting call quality.

\begin{figure}[h]
\centering
\vspace{-9mm}
\includegraphics[scale=0.26]{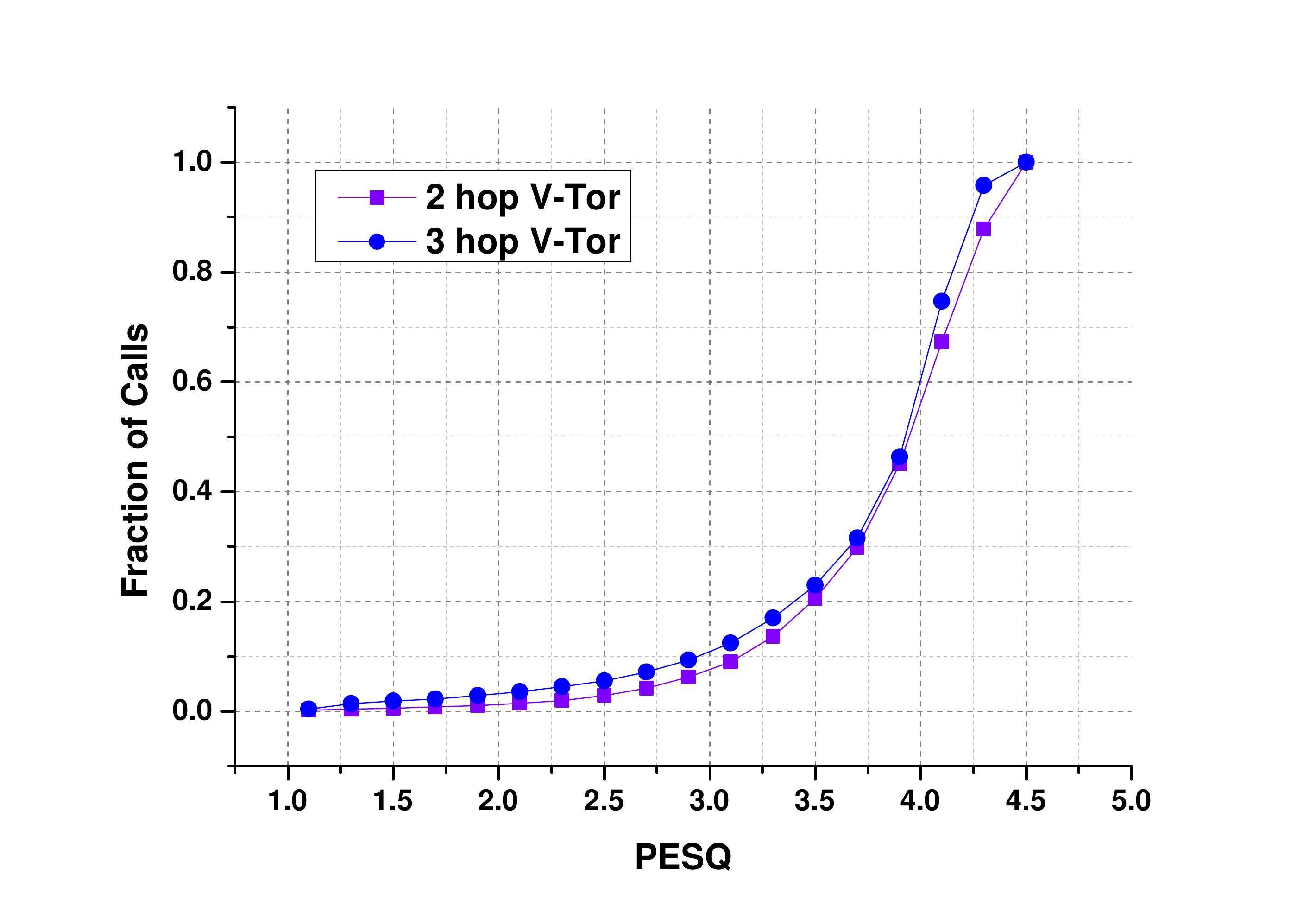}
\vspace{-5mm}
\caption{V-Tor: CDF of PESQ for Caller Anonymity when server is separately hosted (Scenario II)}
\label{vctoc}
\vspace{-6mm}
\end{figure}

To test this, the caller made 1000 individual calls (through Tor)
to every callee (seven locations), via VPN/SIP (or Murmur) servers (three locations).
Similar to the previous experiment, a total of 42000 calls were conducted. 
We observed acceptable quality with average PESQ 3.81 and 
average OWD 270 ms, slightly higher than the previous scenario.

\begin{figure}[h]
\centering
\vspace{-4mm}
\includegraphics[scale=0.26]{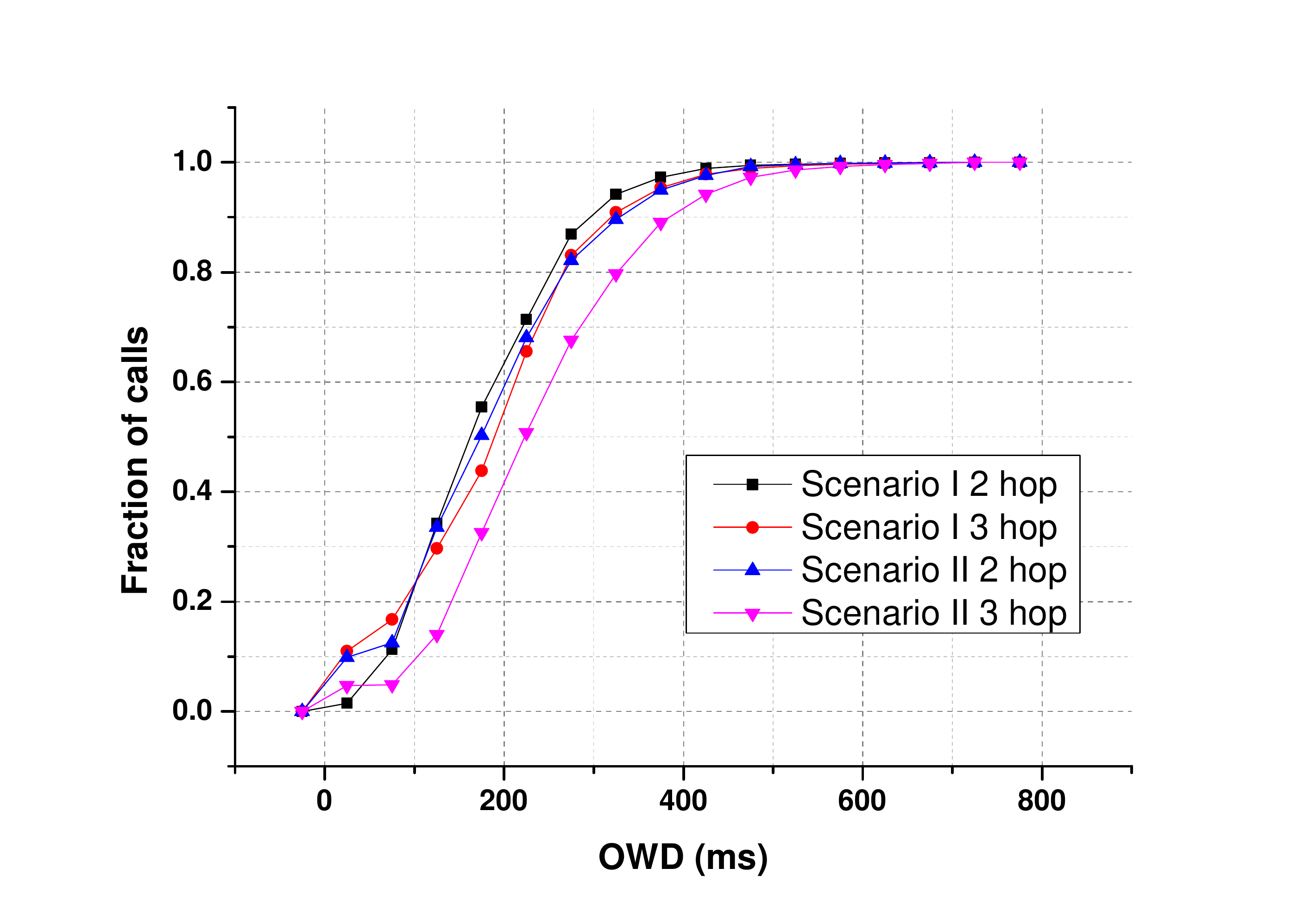}
\vspace{-5mm}
\caption{V-Tor: CDF of OWD variation for Caller Anonymity in both Scenario I and II}
\label{delayvcsc}
\vspace{-5mm}
\end{figure}

Here again, we tried to optimize the performance using shorter 2-hop
circuits. We saw the average PESQ increase to 3.91, and the average OWD reduced to 210 ms.
The results are presented in Fig.~\ref{vctoc}.
A consolidated CDF of OWD for 
V-Tor 
is depicted in Fig.~\ref{delayvcsc}.
Evident from the results, we observed PESQ \textgreater 3 and OWD \textless 400 ms in over 
92\% of the calls across both V-Tor and M-Tor.



\subsubsection{\textbf{Caller and Callee (two-way) anonymity (Scenario III)}}
\label{sec:results:scenario3}
\vspace{-3mm}
There may be cases where both the caller and callee are positioned in
censored networks. In such cases, they may connect via Tor
to a VPN/SIP server placed outside their respective censors' jurisdictions.
Their calls would be routed via their individual Tor circuits.
We thus tried to observe the impact of such scenarios (traffic
traversing two circuits) on the overall call quality as the additional network hops may increase OWD.


Similar to previous experiments, we varied the caller
and callee locations across seven countries, while the VPN/SIP
server was distributed across three. 
Each caller initiated 1000 voice calls to a callee, resulting in a total of 42000 calls. 
For V-Tor we observed an average PESQ of about 3.2 with 81\% calls above PESQ 3.
However, the average OWD, as expected due to the increased network hops, was about 458ms. This is slightly above the acceptable limit. 

\begin{figure}[h]
\centering
\vspace{-1.2cm}
\includegraphics[scale=0.26]{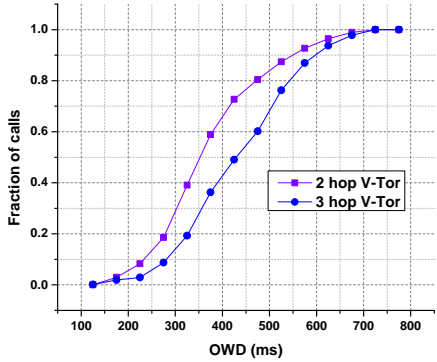}
\vspace{-5mm}
\caption{CDF of delay for V-Tor setup when two-way Anonymity was achieved (Scenario III).}
\label{vbstor}
\vspace{-6mm}
\end{figure}

We thus tried to optimize performance by using shorter
two-hop circuits. We hence repeated the above tests using two-hop circuits
and observed a reduction of the average OWD to 396 ms. The results are shown in Fig.~\ref{vbstor}.
In general, for such scenarios, regular three-hop circuits
incur higher OWD, compared to two-hop ones.
Hence, two-hop circuits seem a better choice for
such cases. 
However, the results of our user study (ref Sec.~\ref{userstudy}) shows that users did not have any noticeable performance impact (due to the delay introduced) when both the users conversed via Tor for two-way anonymity.


\textit{To summarize, in all the three scenarios over the public Tor network
 we observed good call quality (PESQ \textgreater 3 and
OWD \textless 400 ms) in about 85\% cases.
}

We use 400 ms as a threshold for good call quality following the ITU recommendation for international calls~\cite{itu2003recommendation}. However, these recommendations also reported some user dissatisfaction even when OWD was between 300ms and 400ms. On analyzing the results, we found that more than 80\% of calls had OWD $<300ms$. This has been further analyzed in detail in the Appendix.~\ref{app:discussion}.



\vspace{-5mm}
\subsubsection{\textbf{Experiments involving popular apps}}
\vspace{-3mm}
Next, we evaluate the performance of two popularly used VoIP apps, Telegram, and Skype, when running over Tor.
Evaluating these apps would be beneficial from the usability point of view as most users generally use these apps for their day to day tasks. 
Hence using them for anonymous calls would be relatively easy, as they would not require 
installing V-Tor or M-Tor setups.
However, the users might not be completely secure or anonymous, when using these apps (as discussed in Sec.~\ref{sec:Discussion}).

In this experiment, we instantiated 1000 consecutive calls using each of these apps (for both the setups) and computed
their call quality. 
The average PESQ score was 3.8 and 3.54 for Telegram and Skype, respectively. Skype had more than 80\% calls with PESQ score $>$3, whereas Telegram had 85\% $>3$.
Overall, there was not much difference in terms of call quality between the two applications.

\textit{In our results, we observe that popular voice calling apps
perform acceptably well over Tor.} 


\vspace{-4mm}
\subsection{Direct calls}
\vspace{-2mm}

In the previous subsections, we have established that users would in the majority of the cases obtain good call quality when calls are performed over Tor.
However, it would be interesting to compare the performance when calls are instantiated with and without Tor to understand the relative change in call quality. 

\begin{figure}[h]
\vspace{-7mm}
\centering
\includegraphics[scale=0.26]{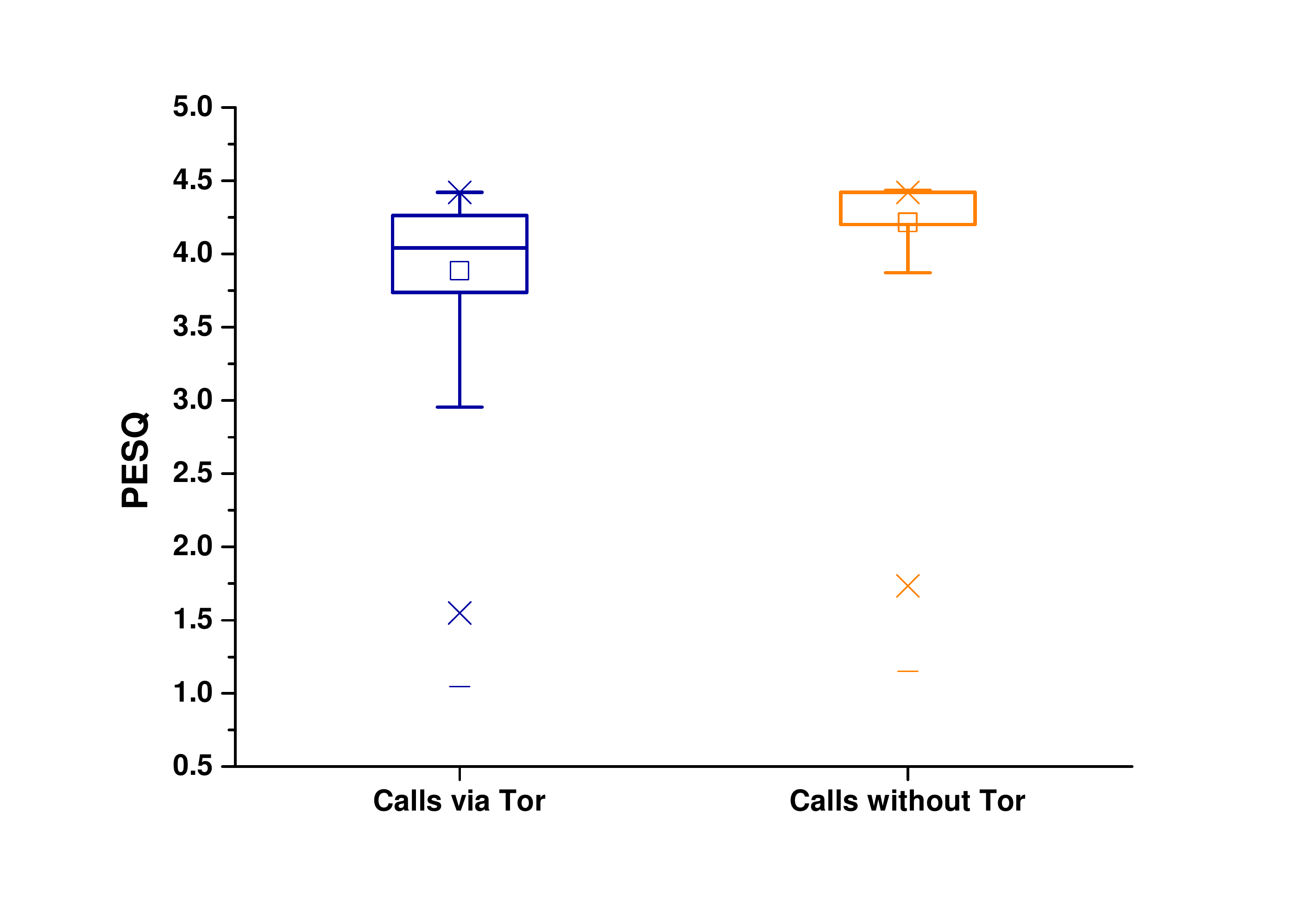}
\vspace{-7mm}
\caption{Comparison of call quality in terms of PESQ obtained with and without Tor.}
\label{comparison}
\vspace{-5mm}
\end{figure}


Hence, we performed direct calls involving both the setups. In the V-Tor setup, the calls were carried via VPN, and for M-Tor they were carried out directly using mumble.  
We instantiated 1000 calls for each scenario described in earlier subsections. The results are shown in Fig.~\ref{comparison}.
As can be seen from the results, there is an expected relative drop in performance when moving from non-Tor to Tor setups. This is obvious as there are expected to be much fewer distortions on non-Tor setups. However, the performance of calls via Tor is good enough for the thresholds of performance metrics used in our study (OWD \textless 400ms and PESQ \textgreater 3).

\vspace{-6mm}
\subsection{Users' Perspective}
\label{userstudy}
\vspace{-3mm}
In our research, we conducted an extensive experimental study to adjudge the quality of voice calls via Tor. The study relied on two standard metrics \emph{i.e.}, PESQ and OWD. These metrics are robust in judging the call quality. In fact, PESQ was established as a substitute for subjective tests (as already discussed in Sec.~\ref{back:Qos}). 

However, in all our experiments, we conducted calls with a maximum duration of 30s. This is because, PESQ does not support the evaluation of calls whose duration is longer than 30s~\cite{itu30s}. Moreover, the PESQ score evaluation is a non-linear function, with respect to call duration. Hence, the mean of multiple samples of 30s will not correspond to the PESQ of the actual combined duration call, as clearly stated by ITU~\cite{itu30s}.
As per ITU recommendations~\cite{itu30s} and PESQ~\cite{pesq} draft, a duration of 8s - 12s is sufficient to judge the call quality of the channel under test. 
However, one might argue that the real users' experience might be different for calls $>30s$ as these durations are not tested for. 
Therefore, we conducted a user study that involved human subjects evaluating the call quality (conducted 
via appropriate ethical review committee's approval, ref.~\ref{app:discussion}).

The user study involved 20
participants\footnote{The location of participants does not have much effect as Tor clients by default select relays in varied locations. However, our participants were spread across three countries.}, which were randomly divided into five groups of four participants each.
The groups were given individual sets of calls, each containing three different recorded calls. 
These calls were conducted over the Tor network, using setups described in Subsection~\ref{setup}

The first call was 30s long, and the remaining were two and four minutes long. 
We calculated PESQ for the 30s call, and recorded OWD and jitter for longer duration calls (as PESQ cannot be calculated for calls longer than 30s).
The users listened to these recorded files and gave an Absolute Category Rating (ACR)~\footnote{This is in accordance with ITU guidelines~\cite{p830itu} for rating calls in subjective tests.} in the range of 1-5.
 We then calculated the Mean Opinion Score (MOS) by averaging the score given by all four participants of each group (results of which are summarized in Tab.~\ref{tab:usermos}). 
 The users reported an average MOS of 4.25, 4.5 and 4.0 for the 30s, 2mins, and 4mins calls, respectively.
 These results show that, in general, users reported good call quality over Tor.
\begin{table}[h]
\vspace{-2mm}
\centering
\small
\begin{adjustbox}{max width=\columnwidth}
\begin{tabular}{c c c c l}
\hline
\textbf{Group} & & \textbf{MOS} \\
\textbf{No.} & \textbf{30s} & \textbf{2min} & \textbf{4min} \\\hline
I & 4.5 & 4.5 & 4 \\\hline
II & 4.5 & 4.25 & 4.5  \\\hline
III & 3.5 & 4.75 & 4.25  \\\hline
IV & 4.25 & 4.5 & 3 \\\hline
V & 4.75 & 4.5 & 4.25 \\\hline
\end{tabular}
\end{adjustbox}
\caption{MOS by different groups for varied call length.}
\label{tab:usermos}
\vspace{-7mm}
\end{table}

\noindent\textbf{Comparing MOS with network attributes:}
We confirmed the aforementioned MOS values (obtained from users) against recorded performance metrics.  
 For the 30s call, we compared the MOS values with the corresponding PESQ scores.
 We observed an average MOS of 4.25, and a correlated average PESQ of 4.2, thus supporting
 our observations. 
 Similarly, for the remaining two calls, we found that OWD and jitter were well within bounds for ``good quality'' (OWD $<400ms$ and jitter $<30ms$). The average OWD was $\approx$278ms, and jitter was about 24ms.



All the above experiments involved the users listening to recorded calls. Thereafter, we went a step ahead, and asked ten users to converse daily via Tor for usual conversations. It must be noted that both the users connected via Tor, simulating Scenario III (two-way anonymity). Users reported a score for each of the calls. Call length varied from 1min to a max of half an hour. This experiment was conducted for about 15 days, and users in the majority of the cases reported good call quality comparable to that achieved via popular VoIP applications rating an average MOS of 4.1.

\textit{The results in this user study further strengthen the claim about obtaining adequate call quality for anonymous calls over Tor.}



\vspace{-6mm}
\section{Insights from Measurements}
\label{analysis}
\vspace{-2mm}
We now present explanation of the aforementioned results, along with other interesting insights we observed from these results.

\subsection{Overall Performance Analysis}
Computing PESQ involves the comparison of the
original audio clip, as played out by the caller, 
with what is recorded at the callee.
Effects of network delays, jitters, and losses, 
reflected in the recorded
audio, are captured by this metric.
Variations in network conditions, like 
increase in contentious cross-traffic, 
leads to an increase in the drops and delays, and
thus negatively impacts
the perceived quality (and thus PESQ).
Besides PESQ, such contention also impacts other
network performance metrics like
OWD, RTT and available bandwidth. 

Our overwhelmingly positive results,
with PESQ \textgreater 3 and OWD\textless 400ms in 85\% cases
are indicative of relatively low
network contentions that can impact
VoIP call quality. VoIP calls
are encoded at low sending rates (\textless 120 Kbps) and thus require low available bandwidth.
Further, in 
$\approx$ 90.6\% of our Tor circuits, we observed adequate available bandwidth ($>1$ Mbps), as reported by \texttt{iperf}. 
About 95\% of these 90.6\% circuits supported calls with acceptable performance. This indicates that --- circuits with sufficient available bandwidth improves the chances of call obtaining good perceptual quality.
This can be further understood 
by analyzing the 90.6\% circuits
where we obtained $>1$ Mbps bandwidth. 
Thus, we tabulate the frequency of these circuits, along with their corresponding PESQ scores. As evident from Tab.~\ref{tab:mode}, for calls where we obtained good perceived quality ($PESQ>3$), the frequency of $>1$ Mbps bandwidth circuits were also very high. 
Similarly, we also observed that for bad quality calls $PESQ<3$, the instances of $>1$ Mbps circuits were relatively low. 


\begin{table}[h]
\centering
\small
\begin{adjustbox}{max width=\textwidth}
\begin{tabular}{c c c c c}
\hline
\textbf{PESQ} & 1-2 & 2-3 & 3-4 & 4-5 \\\hline
\textbf{Frequency} & 6K & 22K & 162K & 259K \\\hline
\end{tabular}
\end{adjustbox}
\caption{Variation of frequency of Tor circuits (\textgreater 1Mbps bandwidth) with PESQ of calls via them.}
\label{tab:mode}
\vspace{-8mm}
\end{table}

\begin{figure}[h]
\vspace{-5mm}
\centering
\includegraphics[scale=0.3]{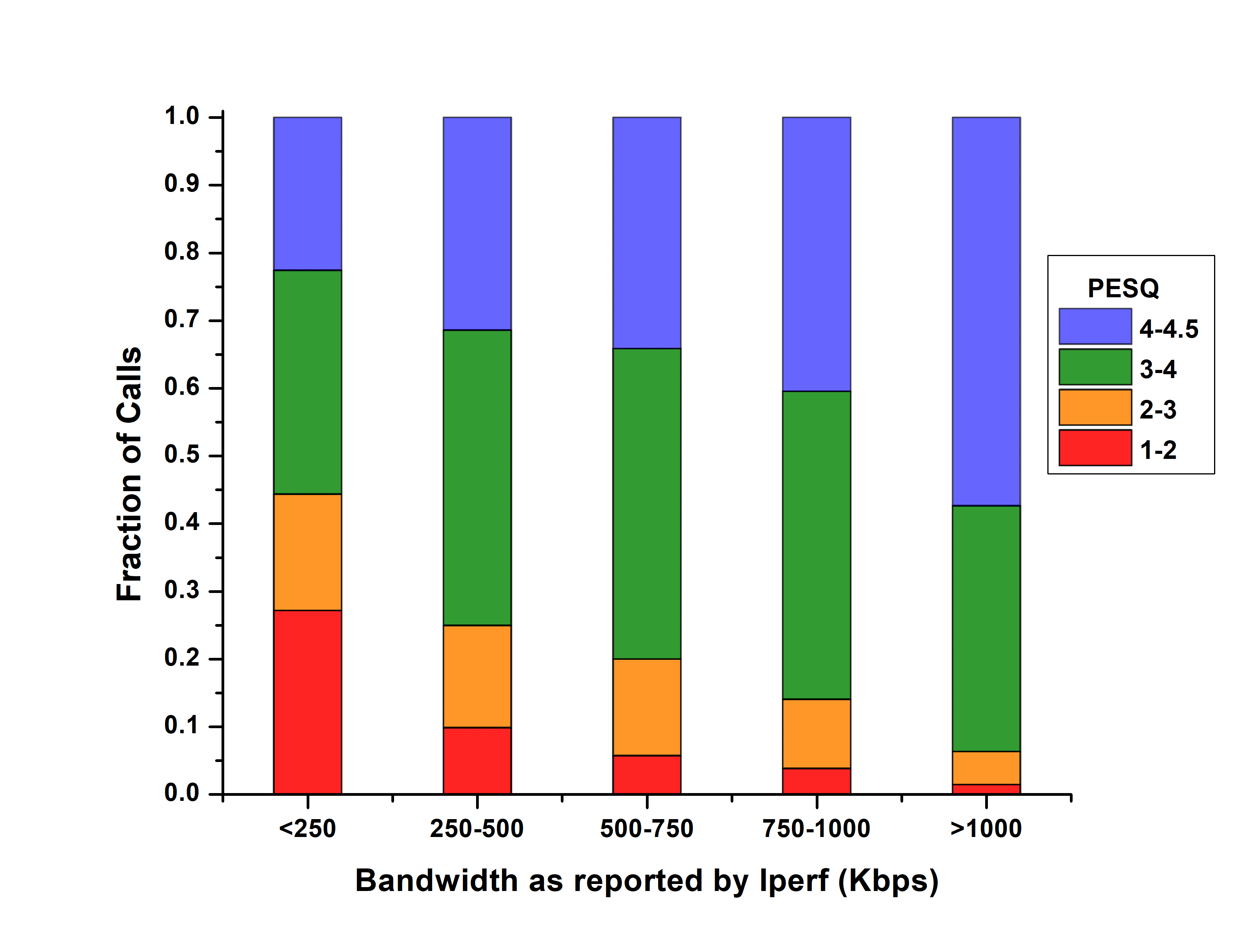}
\vspace{-5mm}
\caption{Fraction of PESQ scores at different available bandwidths.}
\label{BW_frac}
\vspace{-4mm}
\end{figure}

In general, we observe that with an increase in network contentions,
both call quality and available bandwidth decrease. This is
evident from Fig~\ref{BW_frac}. As incidences of high PESQ coincide with cases when the recorded
available bandwidth is high, and vice versa.

\noindent\textbf{Impact On Performance Over Time}
Additionally, we also analyzed whether the performance of VoIP calls changed over time. For this, we randomly sampled $100$ calls from our measurements and analyzed the distribution of calls based on quality. We repeated this experiment several times, and in \textit{each} iteration, we observed that $>85\%$ calls always observed good perceived quality. This indicates that the overall distribution is uniform, and there was no observable change over time.

\vspace{-5mm}

\subsection{Performance Dependence on Types of Relays}
\vspace{-2mm}


\begin{figure}[h]
\vspace{-3mm}
\centering
\includegraphics[scale=0.27]{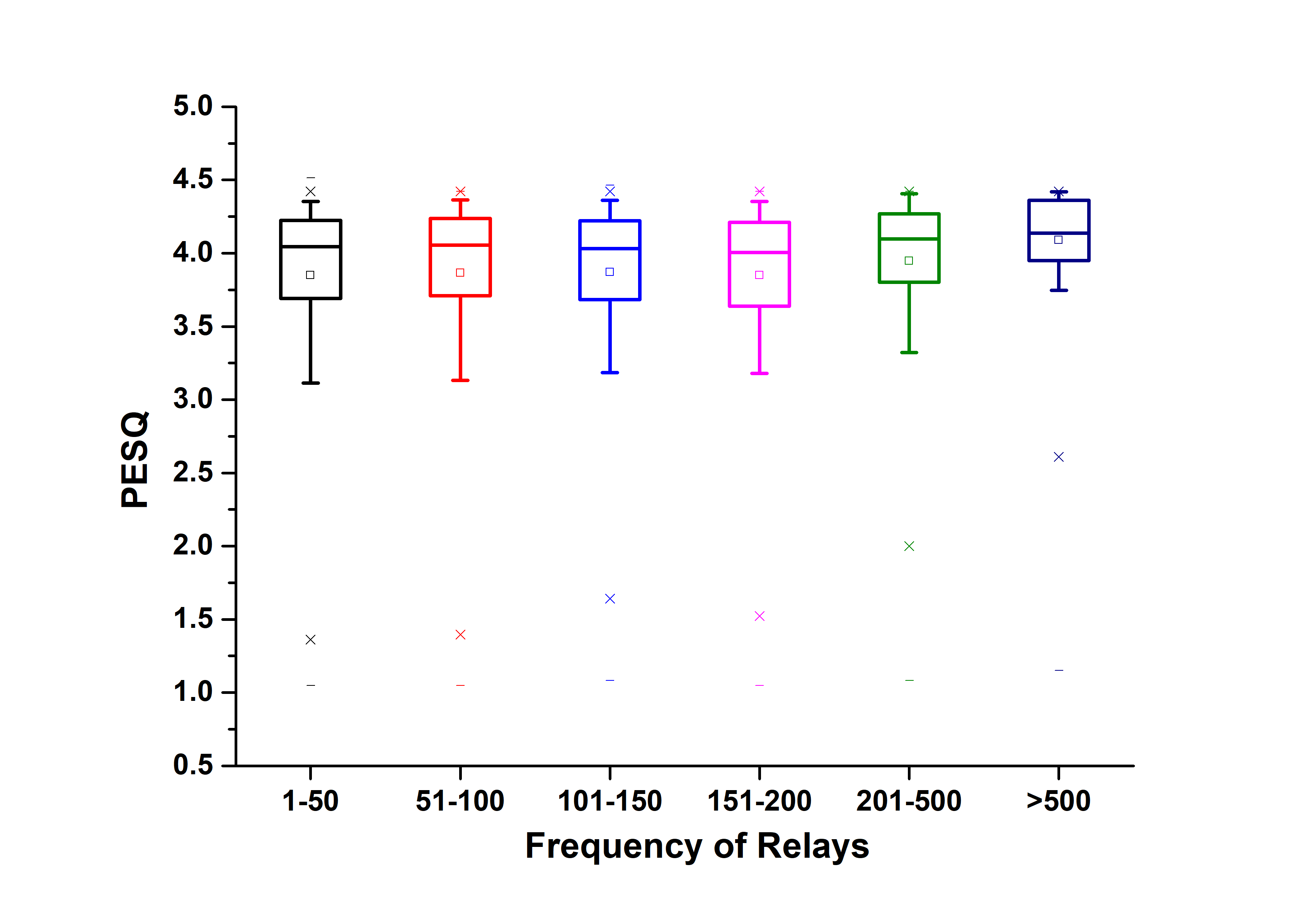}
\vspace{-5mm}
\caption{PESQ of individual calls vs frequency of exit relays.}
\label{pesqfreqall}
\vspace{-4mm}
\end{figure}

In this subsection, we analyzed whether different type of relays (\emph{viz.,} Guard, Middle or Exit) along with their frequencies have any observable impact on the VoIP call quality over Tor.
To begin with, we distributed the relay frequency into three bins --- low ($<10$), moderate ($150-200$) and high ($>500$). 
Then, from each group, we randomly selected a few entry, middle, and exit relays (around ten each) and manually inspected the PESQ scores of the calls involving them.
For each type of relays (in all frequency bins), we observed PESQ scores ranging from 1--4.5, with the majority of them being $>3$. This strongly indicates that PESQ neither depends on any specific type of relay, and nor on their corresponding frequency of occurrence.

\textit{To further ascertain our claims and to obtain a comprehensive picture for all our measurements,
we plotted the distribution of PESQ scores
corresponding to all guard, middle, and exit nodes.} 
The box plot for exit nodes is shown in Fig. \ref{pesqfreqall}. 
As evident, the PESQ values show no dependence on the frequency of occurrence of Tor relays. Corresponding to both less ($<50$) and more frequently appearing relays ($>500$) we observed PESQ $>3$ for a large fraction of calls. The trend was similar for guard and middle relays.

\begin{figure}[h]
\vspace{-0.4cm}
\centering
\includegraphics[scale=0.30]{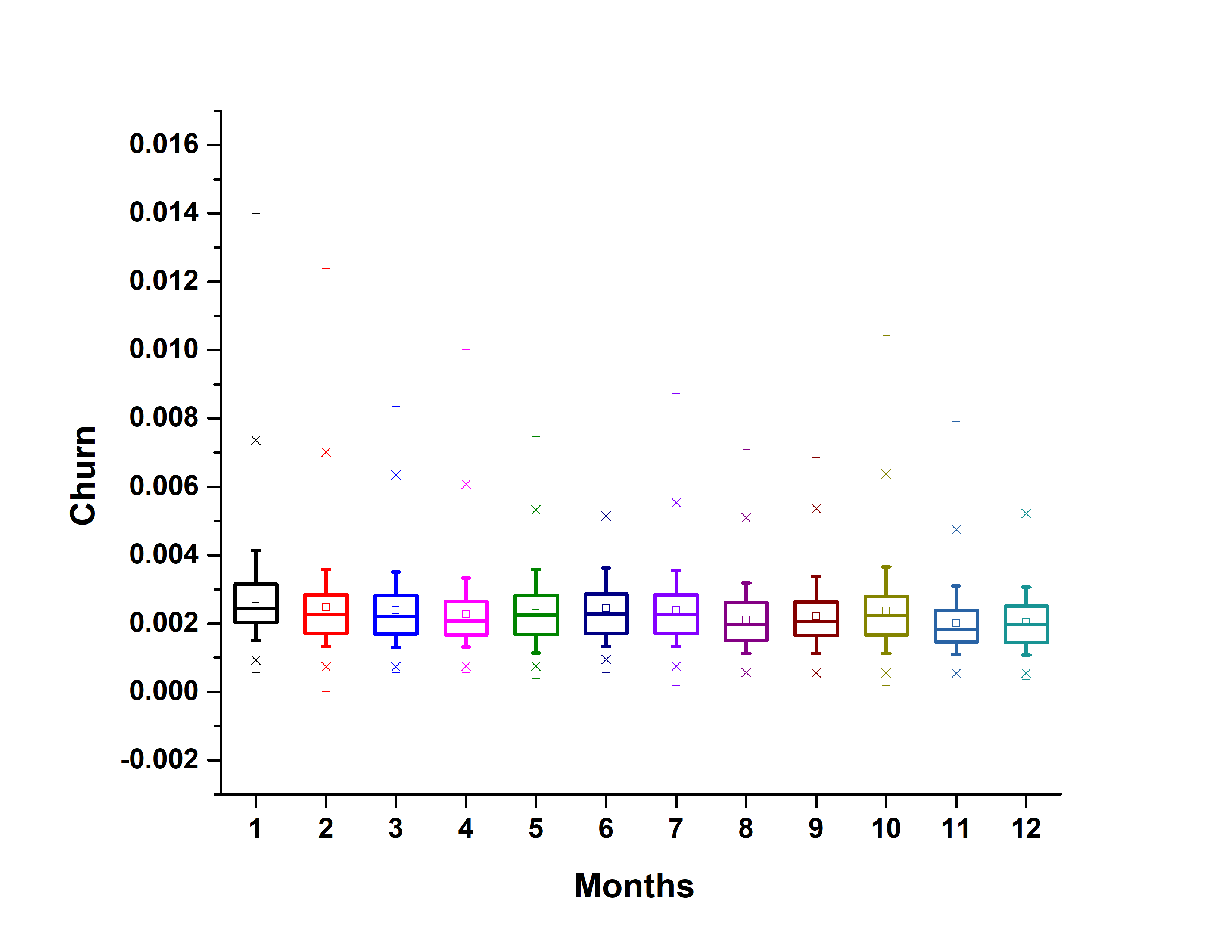}
\vspace{-5mm}
\caption{Tor relay churn for the $12$ months the study was performed.}
\label{churn}
\vspace{-4mm}
\end{figure}

\noindent\textbf{Tor Churn Analysis :}
We now determine whether Tor relay churn had any impact on our results.
Tor relay churn is defined as the rate of relays joining or leaving the network from one consensus to the other (according to Tor Metrics~\cite{tormetrics} and Winter \emph{et al.}~\cite{winter2016identifying}). We calculated the monthly relay churn\footnote{The Tor consensus is updated every hour; in one month it is updated around 720 times. The individual box plot corresponds to the change in consensus of these 720 values. Thus, there are 12 box plots each corresponding to a different month.}
for the entire duration of our study (ref Fig.~\ref{churn}).
It is evident from Fig.~\ref{churn} that relay churn was low (an avg. of $\approx 0.2\%$).
Further, we also observed that
$>85\%$ of our measured VoIP calls
had acceptable quality. 
Our overall results thus indicate that such a low value of churn has an insignificant impact on call quality.

Overall, the analysis in this subsection clearly indicates that the performance was not dependent on any specific type of relay. The apparent independence is because, for a call to be successful, we require less cross traffic contention, and bandwidth of $<120Kbps$ for the entire call duration. The prevalence of such an ecosystem naturally on Tor has already been argued in Sec.~\ref{analysis}.

\vspace{-7mm}

\section{Discussion}
\label{sec:Discussion}
\vspace{-2mm}
In this section, we address concerns like how VoIP performs over Tor 
when using bridges, using different codecs, \emph{etc.}

\noindent\textbf{Call quality over Tor bridges:}
Tor bridges~\cite{torbridges}, are unadvertised entry/guard relays, 
whose information is closely guarded. They are conservatively distributed
either through \emph{BridgeDB}~\cite{bridgedb}, or covertly via out-of-band
means (\emph{e.g.}, emails). Censors may identify
and filter Tor traffic using entry node IP addresses and/or port numbers.
User residing in such networks may use bridges to access other Tor relays
and set up the circuits. 

Hence, to study the impact (if any) when using a bridge, we performed   
1000 measurements, using the V-Tor setup, involving a new Tor circuit each time. 
The average PESQ score was about 3.7.
In about 85\% calls, we observed good performance (PESQ \textgreater 3.0 and OWD \textless 400 ms).
We restricted our measurements to a single bridge node as bridges are scarce.



\begin{figure}[h]
\vspace{-8mm}
\centering
\includegraphics[scale=0.30]{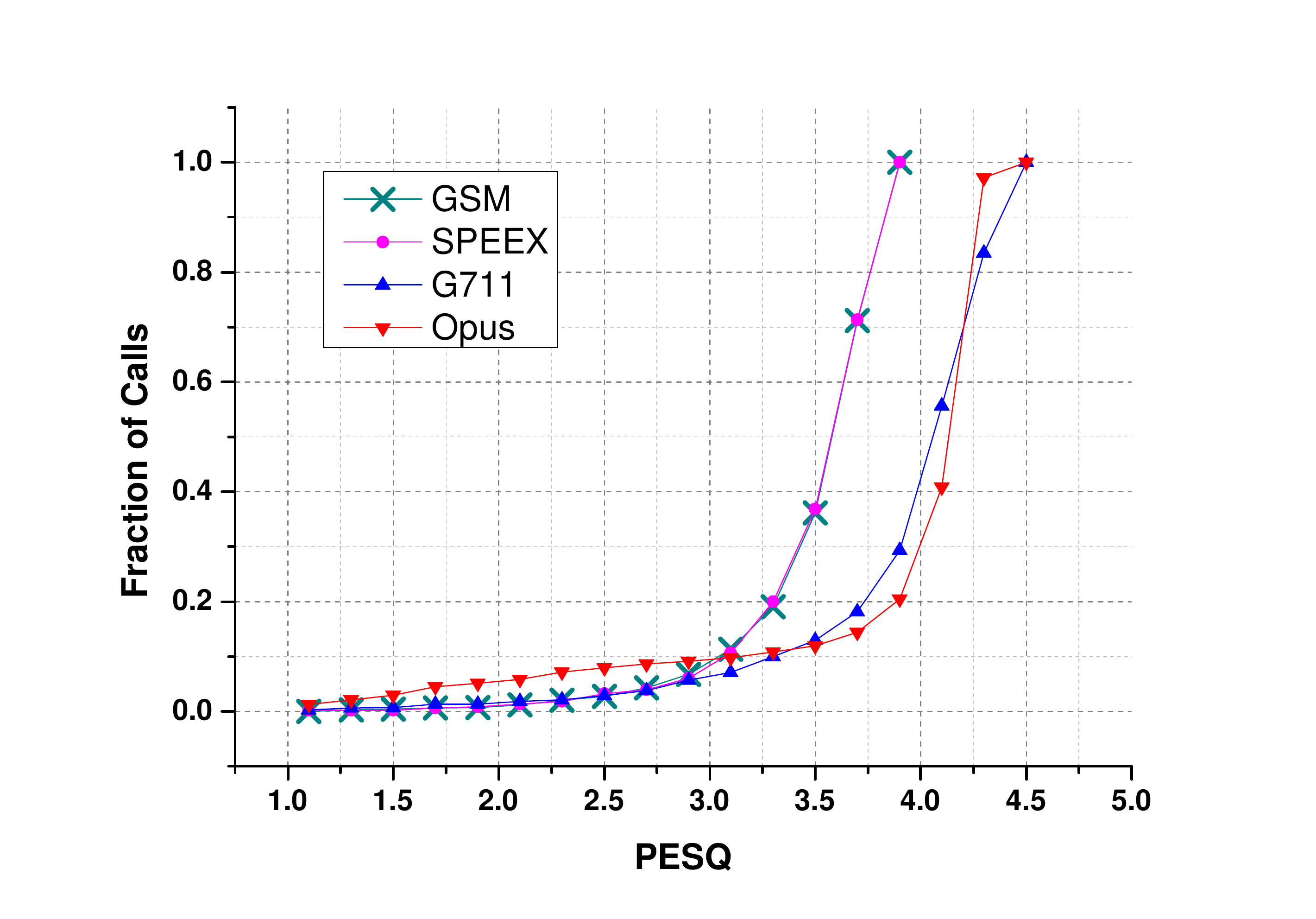}
\vspace{-5mm}
\caption{CDF of PESQ scores when different codecs were used.}
\label{codec}
\vspace{-4mm}
\end{figure}

\noindent\textbf{Impact of codecs:}
Codecs define the way audio is encoded / decoded and transmitted as packets. 
Hence we measured the impact of different codecs on call quality. 
The calls for this experiment were conducted over the real Tor network. We used some popular codecs for our
evaluation, \emph{viz.} \emph{Opus, G.711, Speex, and GSM}. The codecs were selected with the help of configuration changes in the Freeswitch SIP server.
We conducted 1000 calls corresponding to each of these
codecs and measured their PESQ scores (ref. Fig.~\ref{codec}). We observed that the fraction of calls with PESQ \textgreater 3 were roughly equal in all the cases. Thus any of the tested popular codecs could have been used for initiating good quality calls. However, GSM and Speex are lossy codecs, compared to lossless 
codecs like G.711, and thus may not provide high-quality calls (PESQ \textgreater 4)~\cite{ilias2014performance}.
Besides, such lossy codecs encode at lower bitrates compared to the lossless ones \footnote{GSM and Speex encode at bitrates \textless40Kbps whereas G.711 encodes at 84Kbps }. 
Thus, they may be used for Tor circuits with low available bandwidth to receive adequate call quality.

\noindent\textbf{Thresholds for PESQ:}
The MOS scale proposed by ITU delineates perceived performance using 
sharp integral differences. \emph{E.g.}, two corresponds to
``annoying but usable'' and three corresponds to ``fair''.
However, it does not extend such annotations for values in-between.
Further, upon listening to a few calls manually where PESQ was between $2.8$ and $3.2$,
we observed no audible differences. 
Moreover, recent studies~\cite{katsigiannis2018interpreting} 
indicate that humans may be unable to differentiate the
quality of samples, when the difference between their MOS scores is less than 0.4.

In our study, we \emph{conservatively} selected a PESQ of $>$ 3
as ``acceptable'', and anything below it as not (in accordance with ITU). 
We also observed a significant 7\% cases which maybe categorized as ``annoying but usable.'' Finally, we also recorded about 8\% with PESQ under 2. ITU classifies these as ``totally unacceptable.''
Thus, we believe that
actual user experience may be even better than what our study reports (as evident from the conducted user study in Subsec.~\ref{userstudy}).

\noindent\textbf{Selection of caller/callee endpoints:} Throughout our experiments, we used endpoints either as cloud hosts or in-lab machines. All of these were sufficiently provisioned for a voice call with more than adequate bandwidth at the caller/callee end. However, one might argue that our results might be biased as all our endpoints may have sufficient bandwidth for VoIP calls. But, in general, the performance bottleneck was introduced at the Tor relays. As already described in Sec.~\ref{analysis}, there were a significant number of Tor circuits which observed a low bandwidth ($< 1$ Mbps). 
Hence, even if the endpoints are well provisioned, it does not bias our results, as mostly the Tor relays were the bottlenecks.  

\vspace{-5mm}

\section{Conclusion}
\vspace{-1mm}
Real-time anonymous VoIP calls are of interest
to privacy and anonymity conscious citizens,
whistle-blowers, and covert reporters, \emph{etc.}
However, existing research on performance evaluation of VoIP calls over Tor is not comprehensive. They contraindicate 
transporting VoIP packets over Tor and thus favored novel architectures to support anonymous calling. Moreover, there does not exist a functional system to achieve the same. 
Additionally, the costs involved in
recruiting volunteer operated relays (like Tor) along with
users, and managing such a system, 
might outweigh the benefits. 

Thus, it was essential to identify the causes of poor
voice call quality over Tor by observing how the interplay of various
network attributes (RTT, available bandwidth, \emph{etc.})
impacts VoIP quality.
We hence conducted a longitudinal study (spread across 12 months).
It involved extensive testing of about half a million voice calls over Tor, including a user study and using various Tor circuits, peer locations, popular apps, \textit{etc}. 
To our surprise, in over 85\% cases,  we observed acceptably well performance  (PESQ\textgreater 3 and  OWD \textless 400ms),
with only under 8\% cases which were totally unacceptable (PESQ\textless 2). The results of the user study also corroborate
our findings.
Our study is the first to demonstrate that anonymous VoIP calls are indeed possible using Tor.

\section{Acknowledgements}
A special mention to Devashish Gosain, who provided essential comments and feedback throughout the project. He also helped in conducting some experiments.  

This research received no specific grant from any funding agency in the public, commercial, or not-for-profit sectors.

\bibliography{main.bbl}

\begin{thebibliography}{10}

\bibitem{bridgedb}
Tor bridges - bridgedb.
\newblock \url {https://bridges.torproject.org/}.

\bibitem{tor-metrics}
Tor metrics.
\newblock \url {https://metrics.torproject.org/}.

\bibitem{nsaprismskype}
{\em Users guide for PRISM Skype collection}, August 2012.
\newblock \url{https://www.spiegel.de/media/media-35530.pdf}.

\bibitem{nsasurveillance}
{\em NSA uses powerful toolbox in effort to spy on global networks}, December
  2013.
\newblock
  \url{https://www.spiegel.de/international/world/\\the-nsa-uses-powerful-toolbox-in-effort-to-spy-on-global-networks-\\a-940969.html}.

\bibitem{beerends2002perceptual}
{\sc Beerends, J.~G., Hekstra, A.~P., Rix, A.~W., and Hollier, M.~P.}
\newblock Perceptual evaluation of speech quality (pesq) the new itu standard
  for end-to-end speech quality assessment part ii: psychoacoustic model.
\newblock {\em Journal of the Audio Engineering Society 50}, 10 (2002),
  765--778.

\bibitem{cangialosi2015ting}
{\sc Cangialosi, F., Levin, D., and Spring, N.}
\newblock Ting: Measuring and exploiting latencies between all tor nodes.
\newblock In {\em Proceedings of the 2015 Internet Measurement Conference\/}
  (2015), ACM, pp.~289--302.

\bibitem{chaum1981untraceable}
{\sc Chaum, D.~L.}
\newblock Untraceable electronic mail, return addresses, and digital
  pseudonyms.
\newblock {\em Communications of the ACM 24}, 2 (1981), 84--90.

\bibitem{danezis2010drac}
{\sc Danezis, G., Diaz, C., Troncoso, C., and Laurie, B.}
\newblock Drac : An architecture for anonymous low-volume communications.
\newblock In {\em International Symposium on Privacy Enhancing Technologies
  Symposium\/} (2010), Springer, pp.~202--219.

\bibitem{syverson2004tor}
{\sc Dingledine, R., Mathewson, N., and Syverson, P.}
\newblock Tor: The second-generation onion router.
\newblock Tech. rep., Naval Research Lab Washington DC, 2004.

\bibitem{dingledine2009performance}
{\sc Dingledine, R., and Murdoch, S.~J.}
\newblock Performance improvements on tor or, why tor is slow and what we’re
  going to do about it.
\newblock {\em Online: http://www. torproject.
  org/press/presskit/2009-03-11-performance. pdf\/} (2009).

\bibitem{socat}
{\sc Gerhard Rieger}.
\newblock {\em socat}, April 2009.
\newblock \url{http://www.dest-unreach.org/socat/}.

\bibitem{verizon}
{\sc The Guardian}.
\newblock {\em NSA collecting phone records of millions of Verizon customers
  daily}, june 2013.
\newblock
  \url{https://www.theguardian.com/world/2013/jun/06/nsa-phone-records-verizon-court-order}.

\bibitem{sdp}
{\sc Handley, M., Jacobson, V., and Perkins, C.}
\newblock Sdp: session description protocol.
\newblock Tech. rep., 2006.

\bibitem{heuser2017phonion}
{\sc Heuser, S., Reaves, B., Pendyala, P.~K., Carter, H., Dmitrienko, A., Enck,
  W., Kiyavash, N., Sadeghi, A.-R., and Traynor, P.}
\newblock Phonion: Practical protection of metadata in telephony networks.
\newblock {\em Proceedings on Privacy Enhancing Technologies 2017}, 1 (2017),
  170--187.

\bibitem{ilias2014performance}
{\sc Ilias, I. S. H.~C., and Ibrahim, M.~S.}
\newblock Performance analysis of audio video codecs over wi-fi/wimax network.
\newblock In {\em Proceedings of the 8th International Conference on Ubiquitous
  Information Management and Communication\/} (2014), pp.~1--5.

\bibitem{itu2003recommendation}
{\sc ITU-T, I.}
\newblock Recommendation g. 114.
\newblock {\em One-Way Transmission Time, Standard G 114\/} (2003).

\bibitem{itu2000g}
{\sc ITU-T, R., and Recommend, I.}
\newblock G. 114.
\newblock {\em One-way transmission time 18\/} (2000).

\bibitem{onionperf}
{\sc Jansen, R.}
\newblock {\em Onionperf : {A} utility to track Tor and onion service
  performance.}
\newblock The Tor Project, May 2015.
\newblock \url{https://onionperf.torproject.org/onionperf.html}.

\bibitem{jansenshadow}
{\sc Jansen, R., and Hopper, N.}
\newblock Shadow: Running tor in a box for accurate and efficient
  experimentation.
\newblock {\em Proceedings of Network and Distributed Systems Security (NDSS)
  2012\/}.

\bibitem{jansen2019point}
{\sc Jansen, R., Vaidya, T., and Sherr, M.}
\newblock Point break: a study of bandwidth denial-of-service attacks against
  tor.
\newblock In {\em 28th $USENIX$ Security Symposium ($USENIX$ Security 19)\/}
  (2019), pp.~1823--1840.

\bibitem{johnson2017peerflow}
{\sc Johnson, A., Jansen, R., Hopper, N., Segal, A., and Syverson, P.}
\newblock Peerflow: Secure load balancing in tor.
\newblock {\em Proceedings on Privacy Enhancing Technologies 2017}, 2 (2017),
  74--94.

\bibitem{katsigiannis2018interpreting}
{\sc Katsigiannis, S., Scovell, J., Ramzan, N., Janowski, L., Corriveau, P.,
  Saad, M.~A., and Van~Wallendael, G.}
\newblock Interpreting mos scores, when can users see a difference?
  understanding user experience differences for photo quality.
\newblock {\em Quality and User Experience 3}, 1 (2018), 6.

\bibitem{le2015herd}
{\sc Le~Blond, S., Choffnes, D., Caldwell, W., Druschel, P., and Merritt, N.}
\newblock Herd: A scalable, traffic analysis resistant anonymity network for
  voip systems.
\newblock In {\em ACM SIGCOMM Computer Communication Review\/} (2015), vol.~45,
  ACM, pp.~639--652.

\bibitem{mumble}
{\sc Lightspeed gaming LLC}.
\newblock {\em Mumble}, March 2009.
\newblock \url{https://www.mumble.com/}.

\bibitem{mani2018understanding}
{\sc Mani, A., Wilson-Brown, T., Jansen, R., Johnson, A., and Sherr, M.}
\newblock Understanding tor usage with privacy-preserving measurement.
\newblock In {\em Proceedings of the Internet Measurement Conference 2018\/}
  (2018), pp.~175--187.

\bibitem{skype}
{\sc Microsoft}.
\newblock {\em Skype}, August 2003.
\newblock \url{https://skype.com/}.

\bibitem{mplayer}
{\sc The Mplayer Project}.
\newblock {\em Mplayer}, January 2000.
\newblock \url{http://www.mplayerhq.hu/design7/news.html}.

\bibitem{openvpn}
{\sc OpenVPN INC.}
\newblock {\em OpenVPN}, November 2006.
\newblock \url{https://www.openvpn.net/}.

\bibitem{panchenko2008performance}
{\sc Panchenko, A., Pimenidis, L., and Renner, J.}
\newblock Performance analysis of anonymous communication channels provided by
  tor.
\newblock In {\em 2008 Third International Conference on Availability,
  Reliability and Security\/} (2008), IEEE, pp.~221--228.

\bibitem{perry2009torflow}
{\sc Perry, M.}
\newblock Torflow: Tor network analysis.
\newblock {\em Proc. 2nd HotPETs\/} (2009), 1--14.

\bibitem{pfitzmann1991isdn}
{\sc Pfitzmann, A., Pfitzmann, B., and Waidner, M.}
\newblock Isdn-mixes: Untraceable communication with very small bandwidth
  overhead.
\newblock In {\em Kommunikation in verteilten Systemen\/} (1991), Springer,
  pp.~451--463.

\bibitem{pjsua}
{\sc PJSIP}.
\newblock {\em pjsua}.
\newblock \url{https://www.pjsip.org/pjsua.htm}.

\bibitem{pactl}
{\sc Pulseaudio}.
\newblock {\em pactl}, June 2011.
\newblock \url{https://linux.die.net/man/1/pactl}.

\bibitem{p830itu}
{\sc Rec, I.}
\newblock P. 830: Subjective performance assessment of digital telephone-band
  and wideband digital codecs.
\newblock {\em International Telecommunication Union, Geneva (Switzerland)\/}
  (1996).

\bibitem{itu30s}
{\sc Rec, I.}
\newblock P. 862.3: Application guide for objective quality measurement based
  on recommendations p. 862, p. 862.1 and p. 862.2.
\newblock {\em International Telecommunication Union, Geneva\/} (2005).

\bibitem{rix2003comparison}
{\sc Rix, A.~W.}
\newblock Comparison between subjective listening quality and p. 862 pesq
  score.
\newblock {\em Proc. Measurement of Speech and Audio Quality in Networks
  (MESAQIN’03), Prague, Czech Republic\/} (2003).

\bibitem{pesq}
{\sc Rix, A.~W., Beerends, J.~G., Hollier, M.~P., and Hekstra, A.~P.}
\newblock Perceptual evaluation of speech quality (pesq)-a new method for
  speech quality assessment of telephone networks and codecs.
\newblock In {\em Acoustics, Speech, and Signal Processing, 2001.
  Proceedings.(ICASSP'01). 2001 IEEE International Conference on\/} (2001),
  vol.~2, IEEE, pp.~749--752.

\bibitem{rizal2014study}
{\sc Rizal, M.}
\newblock {\em A Study of VoIP performance in anonymous network-The onion
  routing (Tor)}.
\newblock PhD thesis.

\bibitem{sip}
{\sc Rosenberg, J., Schulzrinne, H., Camarillo, G., Johnston, A., Peterson, J.,
  Sparks, R., Handley, M., and Schooler, E.}
\newblock Sip: session initiation protocol.
\newblock Tech. rep., 2002.

\bibitem{schatz2017reducing}
{\sc Schatz, D., Rossberg, M., and Schaefer, G.}
\newblock Reducing call blocking rates for anonymous voice over ip
  communications.
\newblock In {\em Ultra Modern Telecommunications and Control Systems and
  Workshops (ICUMT), 2017 9th International Congress on\/} (2017), IEEE,
  pp.~382--390.

\bibitem{rtp}
{\sc Schulzrinne, H., Casner, S., Frederick, R., and Jacobson, V.}
\newblock Rtp: A transport protocol for real-time applications.
\newblock Tech. rep., 2003.

\bibitem{freeswitch}
{\sc SignalWire}.
\newblock {\em Freeswitch}, January 2006.
\newblock \url{https://freeswitch.com/}.

\bibitem{snader2009eigenspeed}
{\sc Snader, R., and Borisov, N.}
\newblock Eigenspeed: secure peer-to-peer bandwidth evaluation.
\newblock In {\em Proceedings of the 8th international conference on
  Peer-to-peer systems}.

\bibitem{snader2010improving}
{\sc Snader, R., and Borisov, N.}
\newblock Improving security and performance in the tor network through tunable
  path selection.
\newblock {\em IEEE Transactions on Dependable and Secure Computing 8}, 5
  (2010), 728--741.

\bibitem{sun2004speech}
{\sc Sun, L.}
\newblock {\em Speech quality prediction for voice over internet protocol
  networks}.
\newblock PhD thesis, 2004.

\bibitem{libtgvoip}
{\sc Telegram}.
\newblock {\em libtgvoip}, February 2017.
\newblock \url{https://github.com/grishka/libtgvoip}.

\bibitem{pyrogram}
{\sc Telegram}.
\newblock {\em pyrogram}, January 2018.
\newblock \url{https://github.com/pyrogram/pyrogram}.

\bibitem{tg}
{\sc Telegram Messenger LLP}.
\newblock {\em Telegram}, August 2013.
\newblock \url{https://telegram.com/}.

\bibitem{torsocks}
{\sc Tor dev team}.
\newblock {\em Torocks}, December 2000.
\newblock \url{https://linux.die.net/man/8/torsocks}.

\bibitem{torbridges}
{\sc The Tor Project}.
\newblock {\em Tor Bridges}.
\newblock \url{https://2019.www.torproject.org/docs/bridges.html.en}.

\bibitem{tormetrics}
{\sc The Tor Project}.
\newblock {\em The Tor Metrics Project}, January 2009.
\newblock \url{https://metrics.torproject.org/}.

\bibitem{chutney}
{\sc The Tor Project}.
\newblock {\em Chutney.}, February 2011.
\newblock \url{https://github.com/torproject/chutney}.

\bibitem{stem}
{\sc Tor Project}.
\newblock {\em Stem}, March 2013.
\newblock \url{https://stem.torproject.org/}.

\bibitem{torfone}
{\sc Tor Project}.
\newblock {\em Torfone}, April 2013.
\newblock \url{http://torfone.org}.

\bibitem{sbws}
{\sc The Tor Project}.
\newblock {\em Simple {B}andwidth {S}canner.}, March 2018.
\newblock \url{https://github.com/torproject/sbws}.

\bibitem{winter2016identifying}
{\sc Winter, P., Ensafi, R., Loesing, K., and Feamster, N.}
\newblock Identifying and characterizing sybils in the tor network.
\newblock In {\em 25th $USENIX$ Security Symposium ($USENIX$ Security 16)\/}
  (2016), pp.~1169--1185.

\end{thebibliography}







\appendix 
\label{appendix}
\section{Additional discussion points}
\noindent\textbf{VoIP apps with high bandwidth requirement:} 
\label{app:discussion}
Our measurements involved testing VoIP applications 
that encoded at low bit-rates ($<120 Kbps$). However, these apps can be configured to encode at higher rates ($\approx$ 800 Kbps).
We evaluated the performance at these higher encoding rates (200 Kbps, 400 Kbps and 800 Kbps). We ran 1000 calls when the clients
were configured to encode at these rates. 

As expected, an increase in the rates progressively decreased the
measured PESQ. Average PESQ scores were 3.6, 3.2 and 3.0 for
200, 400, and 800 Kbps rates, respectively. However, even at 800 Kbps, we measured PESQ \textgreater 3 for 65\% of the cases. Thus even at higher encoding rates, one can expect reasonable call quality.

\noindent \textbf{Coverage of Tor relays:}
We recorded and analyzed Tor circuit information for all our experiments.
We now present some interesting insights we observed from this analysis.
A total of about 600,000 Tor circuits were created during our study.\footnote{The number of Tor circuits are slightly higher than the total number of experiments as the two way anonymity experiments involve creating two Tor circuits for a single call.}
These circuits involved a total of 6650 unique Tor relays.
Prior research (Rizal \emph{et al.}~\cite{rizal2014study}) reportedly used
only about 298 relays that too restricted to Europe.


\noindent\textbf{Human speech vs. recorded audio:} In our study, we used PESQ for evaluating perceived call quality. PESQ is an objective evaluation metric and requires no human intervention to assess call quality.
However, one may argue that quality assessment through recorded audio
may not be representative of actual human experiences.
Moreover, the impact of speech artifacts like the language used, intonation, pronunciation,
and other peculiarities may not be captured without human intervention. Interestingly, PESQ is proven
to be robust against such artifacts~\cite{sun2004speech,beerends2002perceptual,rix2003comparison}.

\begin{figure}[h]
\centering
\vspace{-4mm}
\includegraphics[scale=0.26]{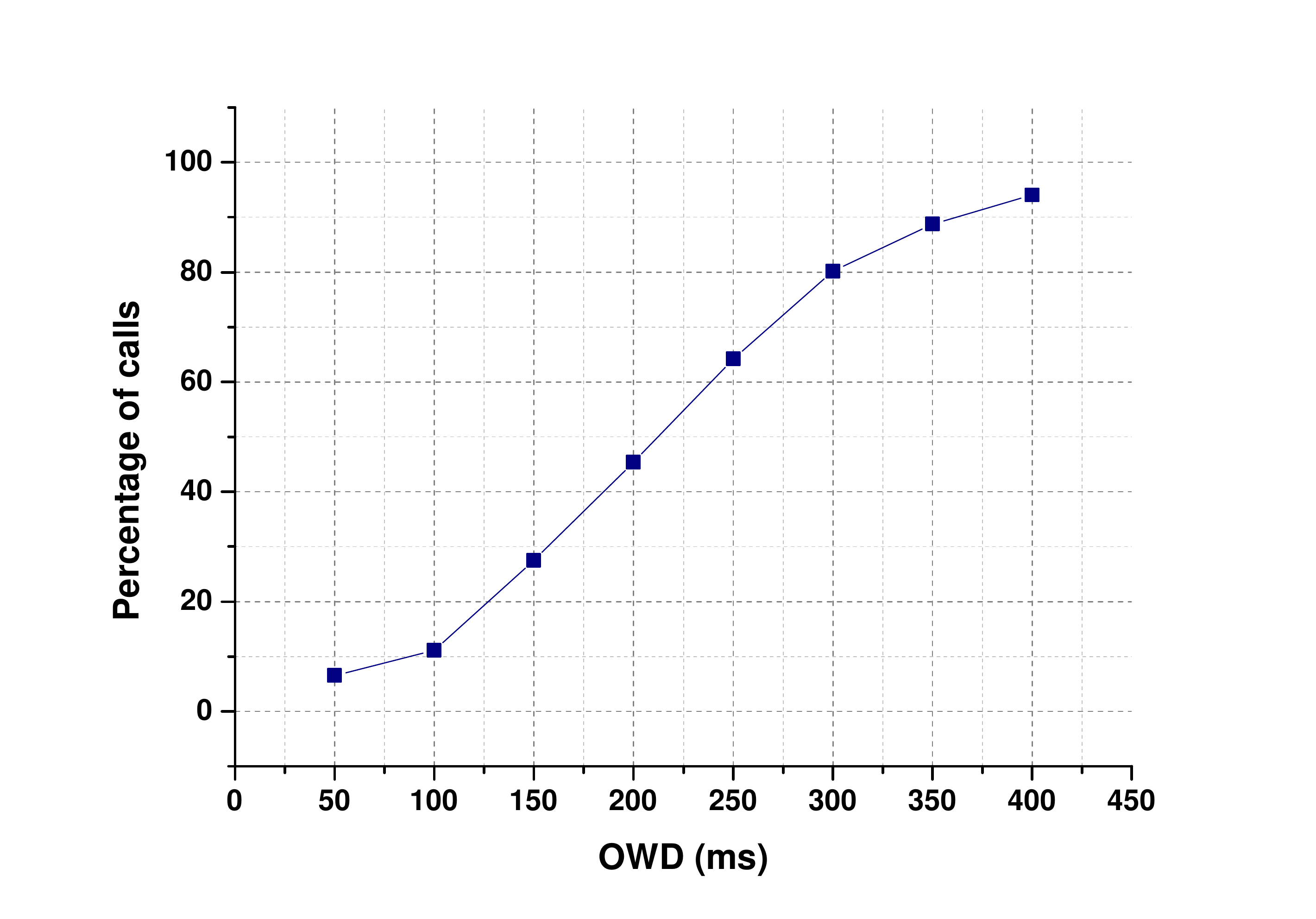}
\vspace{-5mm}
\caption{Percentage of calls recorded at different OWD values.}
\label{owd_overall}
\vspace{-3mm}
\end{figure}

\noindent\textbf{Threshold of OWD:}
Fig.~\ref{owd_overall} shows the percentage of calls which have OWD \textless 50 ms, \textless 100 ms so on till \textless 400 ms.
We notice that only about 10\% of calls had a one-way delay \textgreater 300 ms and \textless 400 ms indicating only a small fraction of calls in that category. On the contrary, for $\approx81\%$ of calls, the OWD was \textless 300 ms. This suggests that in the majority of the cases, the user would obtain satisfactory call quality with OWD \textless 300 ms.
 Moreover, in $\approx$ 28\% of calls OWD was \textless 150 ms, which is regarded as an ideal quality call according to ITU. 
Thus, though we considered the ITU recommendations of
OWD \textless 400 ms to judge call quality, for the majority of our calls, we observed OWD \textless300 ms.

\noindent\textbf{Ethical Considerations:}
As described in the paper, we generated all our network traffic (\textit{i.e.} the voice calls) using machines in our control. We did not capture or use any third party's data/network traffic. Moreover, as our measurements were 
spread across a span of 12 months, and involved generating low bit-rate voice traffic ($\approx$ 120 Kbps),
along with a short duration (\textless 10s) single \texttt{iperf} probes, we expect it to have had negligible to no impact on any un-involved Internet users' network performance.

Our user study involved human subjects in different geographic locations who heard audio samples
and spoke to one another, via our setups. To the best of our knowledge the audio sample bore no 
information that may cause emotional or psychological trauma to the subjects involved. 
The quality and the contents of the clips were duly attested by the institutional 
research review committee that involved subjects who were not party to the research
in any capacity. Further, we did not record and (or) decode, either manually or electronically, the 
speech between subjects, thereby preserving their communication privacy.

\section{M-Tor Results}
\label{app:mtor}
In this section, we describe the implementation details along with the experiments performed using the M-Tor setup. 
The client machine configuration along with how Tor was setup remains the same for M-tor as well (described in detail in Subsec.~\ref{implementation}).
The detail of the new entity introduced in this setup is described below:

\noindent \textit{\textbf{Mumble Server:}} The M-Tor setup had the Mumble server (Murmur) v1.2.19 installed for handling voice calls. A call channel was opened for every new call. The caller and callee were configured to join this call channel so that whenever a caller initiated a call, it
would reach the channel and the callee could record it for quality evaluation.

Now, we describe the experiments involving the controlled setups as well the ones performed over the public Tor network. All these experiments followed the setups similar to those of V-Tor.

\subsection{Controlled Experiments}
Similar to the V-Tor experiments, we performed two different sets of tests 
using M-Tor in the lab environment. 
These involved: (1) Direct Mumble calls without Tor (2) Mumble calls through Tor.
We recorded an average PESQ (measured across 100 individual calls) of 4.5 for both direct calls and calls through Tor. Here also we observed a slightly higher bandwidth requirement ($70-80$ Kbps) for calls that were transported via Tor,
compared to those that were not ($50-60$ Kbps). 
In general, the bandwidth requirement for M-Tor was much lower in comparison to V-Tor as the underlying codec used by Mumble encodes at a lower rate. 

Further, similar experiments were performed where we increased the number of parallel connections gradually, to see its impact on call quality. Here also we observed a trend identical to V-Tor experiments.

\subsection{Experiments over Public Tor}
\begin{figure}[h]
\centering
\vspace{-5mm}
\includegraphics[scale=0.24]{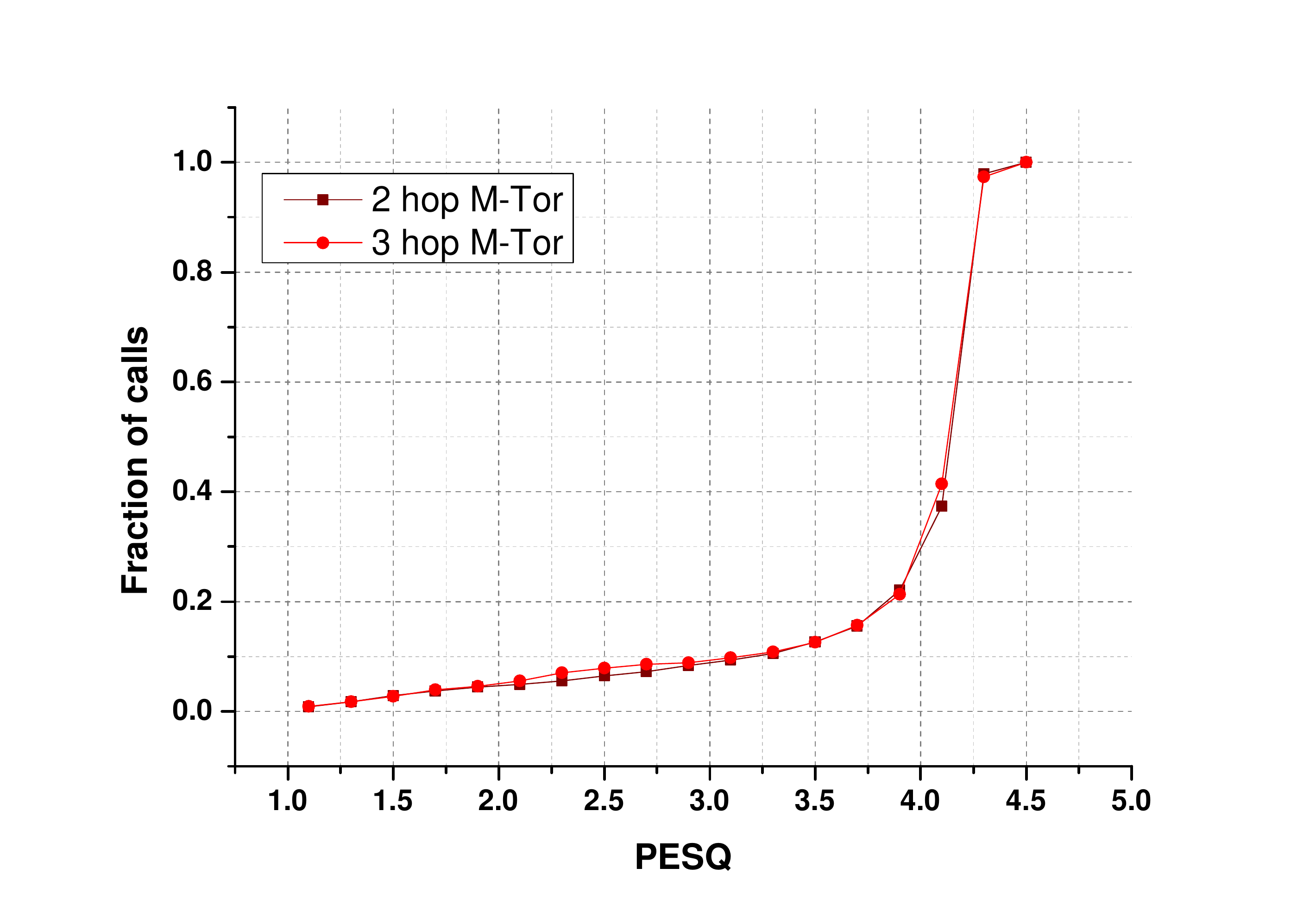}
\vspace{-5mm}
\caption{M-Tor: CDF of PESQ for Caller Anonymity when server is co-located with callee (Scenario I).}
\label{mctos}
\vspace{-5mm}
\end{figure}

\begin{figure}[h]
\centering
\includegraphics[scale=0.24]{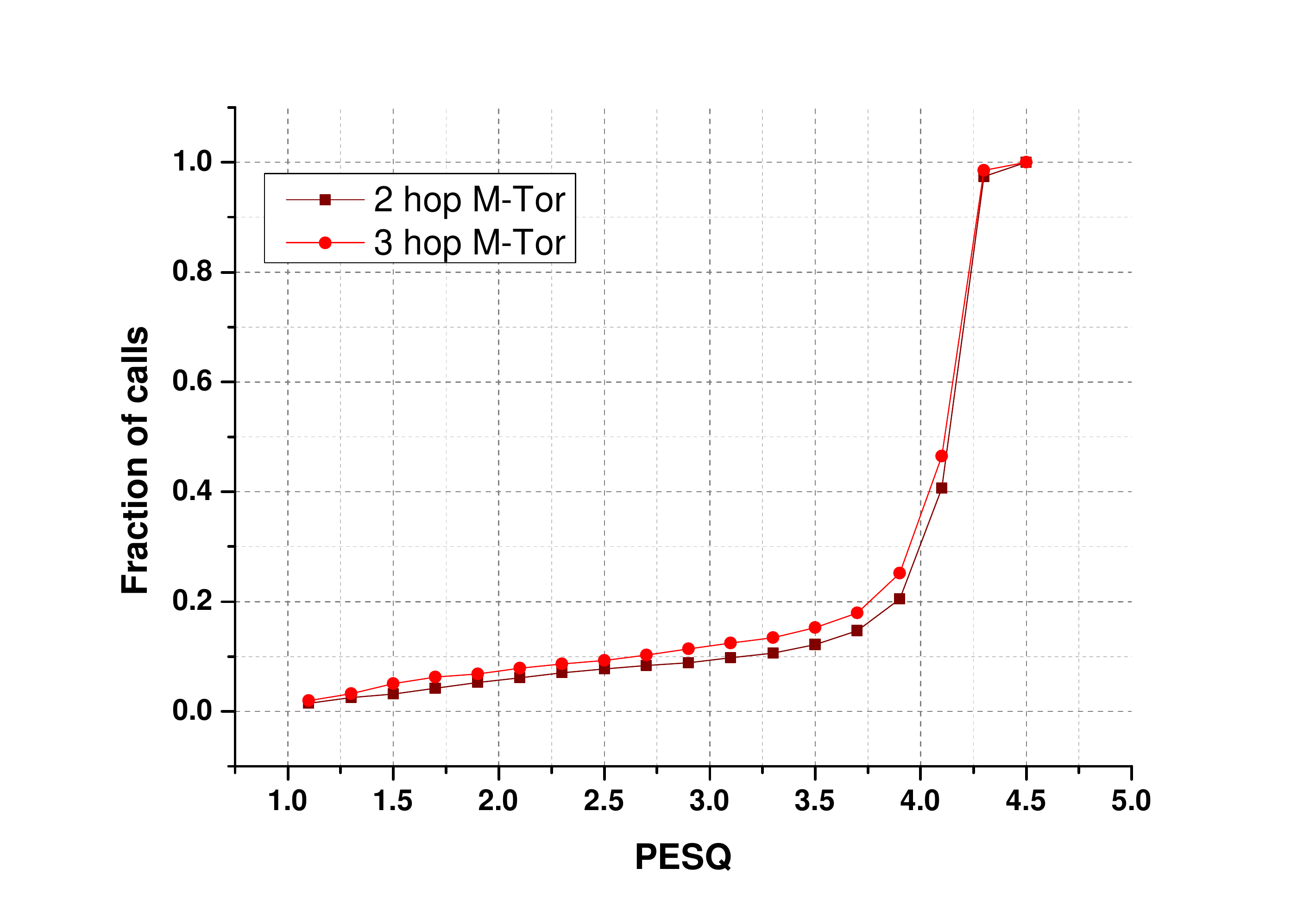}
\vspace{-5mm}
\caption{M-Tor: CDF of PESQ for Caller Anonymity when server is separately hosted (Scenario II)}
\label{mctoc}
\vspace{-5mm}
\end{figure}

We considered the same three scenarios (as described in Subsection.~\ref{sec:results:scenario1}, Subsection.~\ref{sec:results:scenario2} and Subsection.~\ref{sec:results:scenario3}) to perform experiments using M-Tor. 
The CDF of PESQ scores obtained in Scenario I and Scenario II are depicted in Fig.~\ref{mctos}, and Fig.~\ref{mctoc}
respectively.
The average PESQ score obtained in Scenario III, was 3.8 with 85\% calls above PESQ 3.

\begin{figure}[h]
\centering
\includegraphics[scale=0.24]{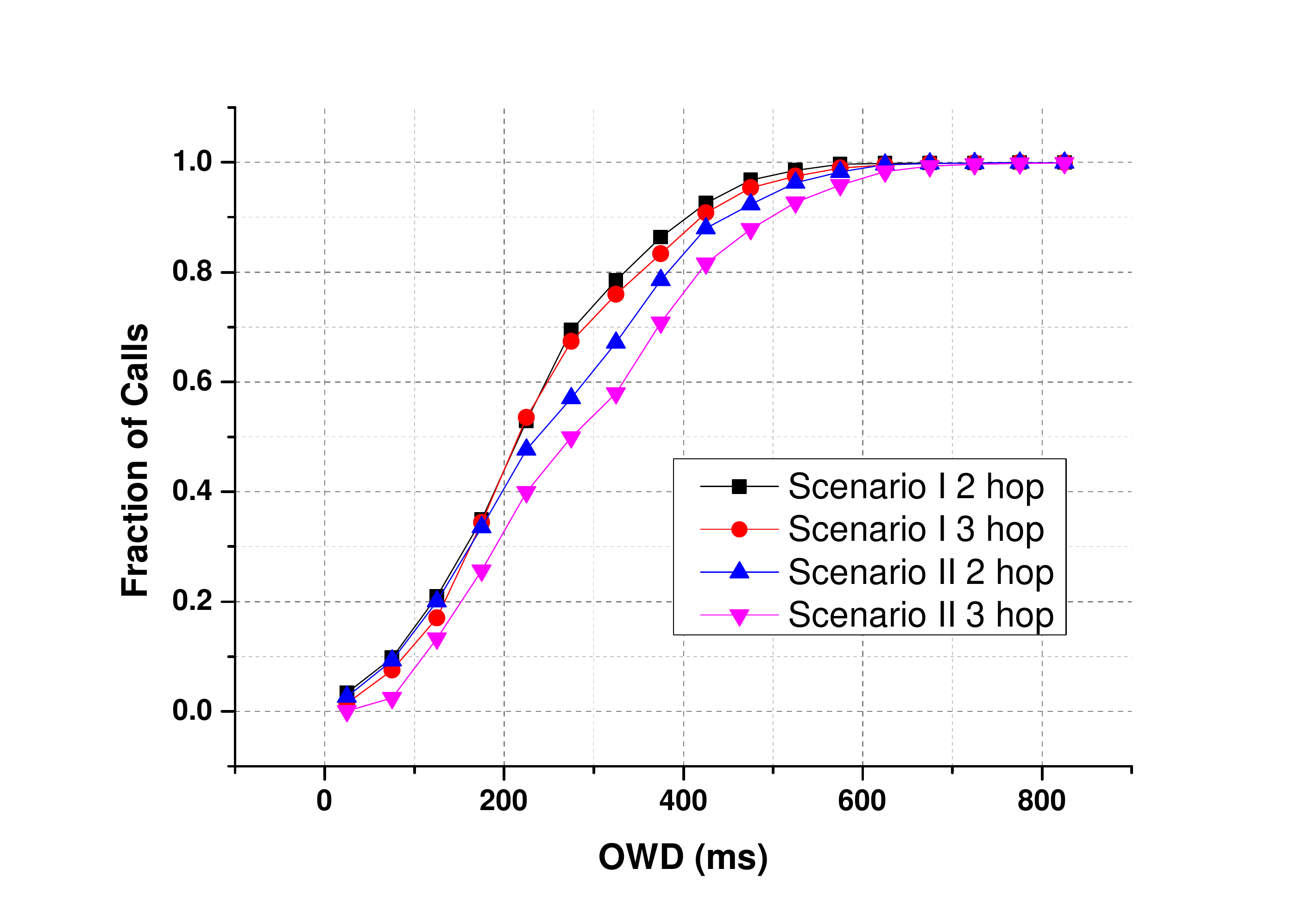}
\vspace{-5mm}
\caption{M-Tor: CDF of OWD variation for Caller anonymity in both Scenario I and II}
\label{delaymcsc}
\vspace{-7mm}
\end{figure}

\begin{figure}[h]
\centering
\includegraphics[scale=0.24]{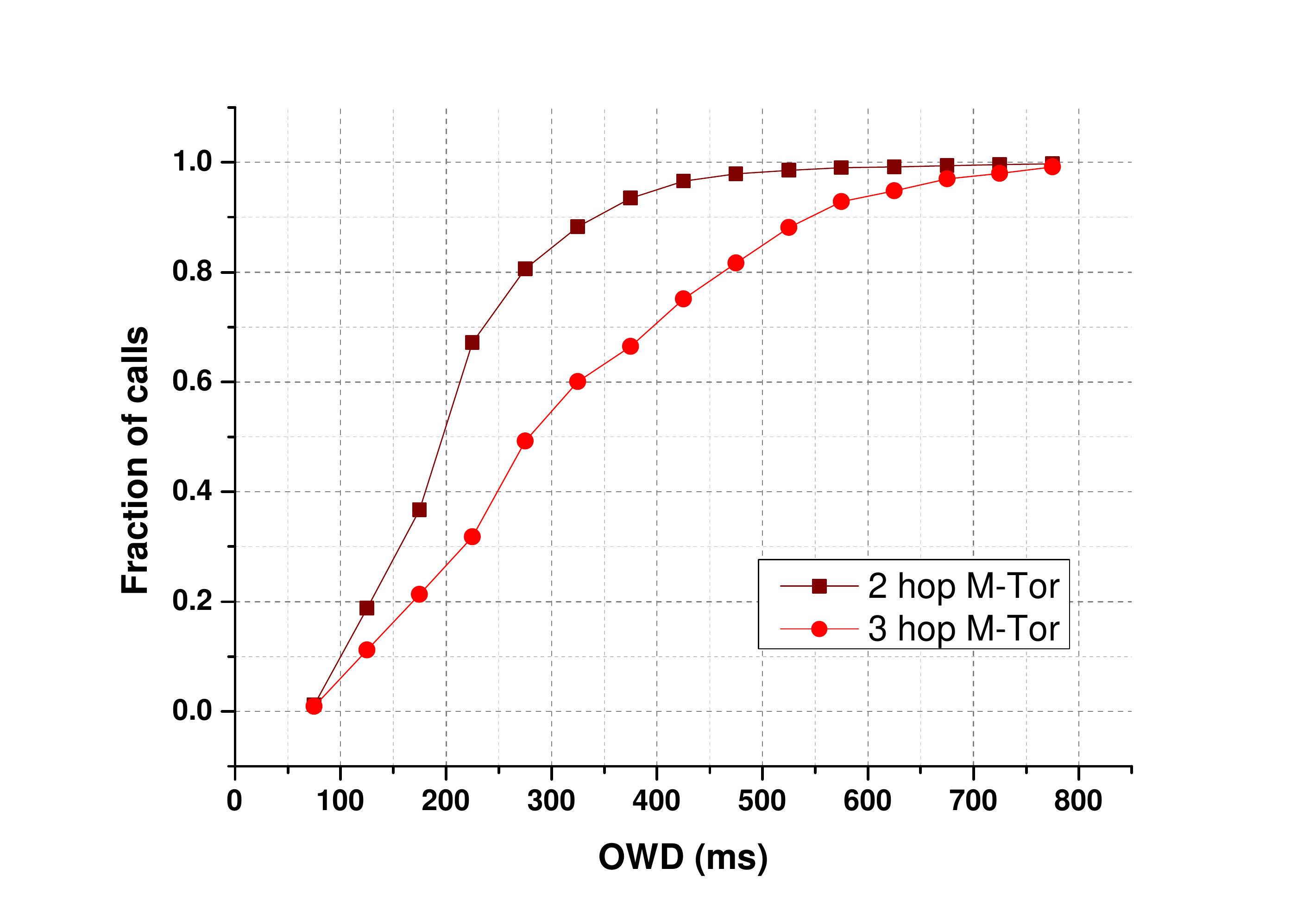}
\vspace{-5mm}
\caption{CDF of delay for M-Tor setup when two-way anonymity was achieved (Scenario III).}
\label{mbstor}
\vspace{-5mm}
\end{figure}

The OWD results for both Scenario I and II (for three hop as well as two hop circuits) are shown in Fig.~\ref{delaymcsc}, with the results of Scenario III in Fig.~\ref{mbstor}. As evident from the results, moving to two-hop circuits in Scenario III (two-way anonymity) helped us improve the OWD significantly with more than 90\% calls below 400ms in comparison to 70\%.

Overall, similar to V-Tor, M-Tor proved to be capable of performing good quality calls for the majority of the cases.

\end{document}